\documentclass[5p,authoryear]{elsarticle} %1p; 3p,twocolumen; 5p; preprint; review; authoryear

\usepackage{framed,multirow}
\usepackage{amssymb}
\usepackage{latexsym}
\usepackage{xcolor}
\usepackage{lineno,hyperref}
\modulolinenumbers[5]
\usepackage{amsmath}
\usepackage{multirow}
\usepackage{color,graphicx}
\usepackage{nicematrix}
\usepackage{arydshln}
\usepackage{hyperref}
\usepackage{url}
\usepackage{footnote}
\usepackage{bm}
\usepackage{cite}
\usepackage{subfigure}
\usepackage{verbatim}
\usepackage{ulem}
\usepackage{graphicx}
\usepackage{pgfplots}
\usepackage{cancel}
\usepackage{dsfont}
\usepackage{adjustbox}

\usepackage{booktabs}  
\usepackage{tabularx}

\usepackage{enumitem}

\journal{Medical Image Analysis}

\begin{document}

\begin{frontmatter}

\title{PhysCARD: Physics-Grounded Reaction-Diffusion Learning for Long-Horizon Electrophysiology Simulation in Cardiac Digital Twins}

\author[label1]{Anh Vu Tran} % anhvutran@u.nus.edu
\author[label1]{Ziheng Deng} % zihengdeng@nus.edu.sg
\author[label1]{Yilin Lyu} % yilinlyu@u.nus.edu
\author[label1,label2]{Mengxiao Wang} % mengxiaowang@sjtu.edu.cn
\author[label3]{Vivek Muthurangu} % v.muthurangu@ucl.ac.uk
\author[label4,label5]{Mark YY Chan} % mark.chan@nus.edu.sg
\author[label4,label5]{Ching-Hui Sia} % mdcschi@nus.edu.sg 
\author[label1]{Lei Li*}

\address[label1]{Department of Biomedical Engineering, National University of Singapore, Singapore}
\address[label2]{School of Mechanical Engineering, Shanghai Jiao Tong University, Shanghai, China}
\address[label3]{Institute of Cardiovascular Science, University College London, London, United Kingdom}
\address[label4]{Department of Medicine, National University of Singapore, Singapore}
\address[label5]{Department of Cardiology, National University Heart Centre Singapore, Singapore}

\cortext[cor1]{Corresponding author: Lei Li (\href{mailto:lei.li@nus.edu.sg}{lei.li@nus.edu.sg})}

\begin{abstract}

Cardiac digital twins require efficient, patient-specific electrophysiology (EP) simulations, but conventional numerical solvers are computationally expensive for repeated evaluations. Neural surrogates offer a promising alternative, yet mainly predicting activation-time maps only or operating on simplified geometries. In this study, we propose PhysCARD, a physics-grounded reaction-diffusion graph network for efficient long-horizon EP simulation. Following operator splitting, PhysCARD learns cellular excitation and recovery through a rest-anchored reaction module and spatial electrical propagation through a physics-encoded diffusion module, preserving key physical properties including the resting equilibrium, balanced electrical fluxes, and scar insulation. A staged learning strategy further improves the accuracy and stability of coupled reaction-diffusion dynamics. Experiments on MRI-derived biventricular meshes from post-infarction patients demonstrated that physics-informed loss regularization reduced the rollout transmembrane-potential RMSE of a graph network from 20.9 to 16.1\,mV, while PhysCARD further reduced it to 4.52\,mV. PhysCARD achieved activation- and repolarization-time errors of 1.23 and 3.71\,ms, respectively, without spurious re-excitation. It also generalized to an unseen mesh resolution exceeding one million nodes and achieved up to 168-fold acceleration over openCARP. These results demonstrate that incorporating physical constraints into neural network architectures enables accurate, efficient, and physically consistent EP simulation in digital twins. The code will be released publicly upon acceptance of the manuscript for publication.

\end{abstract}


\end{frontmatter}

% =========================================================
\section{Introduction}
\label{sec:introduction}
% =========================================================

Cardiac digital twins (CDTs) enable personalized assessment of cardiac function and potential treatments~\citep{journal/EHJ/corral2020,journal/NEJM/chrispin2026}, with cardiac electrophysiology (EP) playing a critical role in arrhythmia applications~\citep{journal/NC/arevalo2016,journal/RiBME/li2024}. However, direct measurement of intracardiac electrical activity requires invasive electroanatomical mapping, while non-invasive ECG provides only indirect information with limited spatial resolution~\citep{journal/CM/vicente2026,journal/NHJ/cluitmans2015,journal/FiP/kalinin2019}. Computational EP simulation addresses these limitations by modelling electrical excitation, propagation, and recovery on patient-specific cardiac meshes using governing partial differential equations (PDEs)~\citep{journal/ABME/crozier2016,journal/SR/wang2019,journal/physiol/vincent2015}. Despite their physiological fidelity, conventional numerical solvers remain computationally expensive, particularly for high-resolution 3D geometries and repeated simulations across pacing conditions and tissue parameters~\citep{fresca2020deep,journal/CAM/roy2020}. This computational burden limits their scalability for patient-specific CDTs and large virtual cohorts~\citep{journal/niederer2020}.

Recent advances in deep learning have enabled neural surrogates that approximate cardiac EP dynamics from anatomical information, physiological parameters, and preceding electrical states, substantially reducing computational costs compared with conventional numerical solvers~\citep{journal/RiBME/li2024,fresca2020deep,fresca2021pod}. 
However, many existing approaches have been developed and validated on simplified settings, such as 2D tissue patches~\citep{herrero2022eppinn}, idealized 3D geometries~\citep{chiu2024physics,conf/ICFIM/morier2025lagata}, and restricted anatomical cohorts~\citep{naghavi2026rapid}. Others rely on geometry-specific representations that may limit generalization to unseen patients or pathological conditions~\citep{martinez2026full}. These limitations hinder their application to CDTs, which require reliable EP simulation across patient-specific anatomies, irregular meshes of varying resolutions, scar distributions, and stimulation conditions.

Long-horizon simulation poses a further challenge for autoregressive EP surrogates in which errors in the predicted states are fed back into the model and accumulate over successive steps, altering wavefront propagation and recovery \citep{conf/MICCAI/tran2026,journal/brandstetter2022}.
Physics-informed neural networks (PINNs) \citep{journal/raissi2019} counter this by incorporating the governing equations into learning, and in cardiac EP they have supported activation-map and transmembrane-potential reconstruction \citep{journal/FiP/sahli2020,journal/IEEE/xie2022,lydon2025} and parameter estimation \citep{herrero2022eppinn,conf/MICCAI/toloubidokhti2025}. However, they impose the physics only as a soft loss penalty, which encourages but does not guarantee physical behaviour and is difficult to optimize for reaction-diffusion dynamics \citep{journal/ANIPS/krishnapriyan2021}. Our preliminary study showed that such a penalty stabilizes graph-based rollouts on 3D cardiac meshes \citep{conf/MICCAI/tran2026}. The network, however, still had to infer recovery and anatomy-dependent conduction, including the effects of fibers and scar, from data, which left repolarization prone to drift and spurious re-excitation.

To address these challenges, we propose PhysCARD, a physics-grounded reaction-diffusion graph network for rapid prediction of transmembrane-potential dynamics on irregular cardiac meshes. Instead of learning the state transition as a black box, PhysCARD follows the operator splitting of the monodomain equation and alternates a learned reaction module for cellular kinetics with a learned diffusion module for anatomy-dependent conduction. Applied autoregressively, it predicts the action potential at every node from activation through repolarization. Our main contributions are as follows:
\begin{enumerate}[label=\roman*.]
    \item We propose an operator-split neural simulator that learns both the reaction and the diffusion operators of cardiac EP and runs on patient-specific meshes of different resolutions without retraining.
    \item We design a rest-anchored reaction flow that keeps the resting state an exact equilibrium, and a diffusion operator with balanced pairwise fluxes and electrical insulation at scar boundaries.
    \item We introduce a staged learning strategy that pretrains the reaction module on the cell model, learns conduction with a physics teacher, and fine-tunes both modules jointly.
    \item We conduct extensive experiments on MRI-derived biventricular meshes with scar, evaluating state accuracy, EP fidelity, physical consistency, and generalization to an unseen mesh resolution.
\end{enumerate}

This study substantially extends our preliminary conference work, PI-GNN~\citep{conf/MICCAI/tran2026}, which incorporated reaction-diffusion physics loss in a recurrent graph network. 
While PI-GNN imposed the physics only through its loss, PhysCARD embeds it in the architecture, replacing the monolithic network with separate reaction and diffusion operators and the generic decoder with a physics-encoded conductance head. 
Moreover, PhysCARD explicitly accounts for myocardial scar, modelling scar tissue as non-excitable and electrically insulating to better represent pathological cardiac EP.
The new introduced staged learning strategy further improves long-horizon stability.
Beyond these methodological advances, we substantially expand the experimental evaluation to different mesh resolutions, including an unseen resolution, with comprehensive analyses of prediction accuracy, EP fidelity, physical consistency, computational efficiency, and individual component contributions.

% ===============================
\section{Related Work}
\label{sec:related_work}
% =========================================================

\subsection{Neural Surrogate Models for Cardiac Electrophysiology}
\label{sec:related_surrogates}

The computational cost of cardiac EP simulation has motivated neural surrogates that approximate electrical dynamics without repeatedly solving the governing equations. Early studies explored reduced-order modelling using autoencoders and Proper Orthogonal Decomposition~\citep{fresca2020deep,fresca2021pod}, learned temporal integration through EP-Net and EP-Net 2.0~\citep{ayed2019ep,kashtanova2021ep}, and recurrent networks for efficient EP prediction~\citep{dong2023deep}. However, these approaches were primarily developed on simplified geometries, limiting their applicability to realistic patient-specific cardiac meshes. More recent work has addressed anatomical variability through explicit geometric representations, including point-cloud variational autoencoders~\citep{workshop/li2022} and coordinate-based neural models~\citep{martinez2026full}. In parallel, geometry-independent operator learning has enabled predictions across varying cardiac anatomies: DIMON maps geometries to a common reference domain~\citep{arxiv/yin2024}, while other approaches employ intrinsic manifold coordinates~\citep{zhou2025unified}, Fourier and kernel operators~\citep{ziarelli2026learning}, or geometry-conditioned latent grids~\citep{journal/ANIPS/li2023gino,naghavi2026rapid}. Although these methods improve anatomical generalization, several focus on activation or repolarization maps rather than complete spatiotemporal EP dynamics. Moreover, their validation on clinically derived biventricular meshes with myocardial scar remains limited.
Most fundamentally, they generally remain fully data-driven, so the governing EP equations are never enforced during training, which limits physical interpretability.

\subsection{Physics-Informed Learning for Cardiac Electrophysiology} \label{sec:related_physics_informed}

To improve the data efficiency and physical consistency of neural surrogates, PINNs \citep{journal/raissi2019} incorporate governing PDEs into the training objective. In cardiac EP, PINNs have been employed to reconstruct activation maps using the Eikonal equation \citep{journal/FiP/sahli2020}, and to recover action potentials and estimate tissue parameters using the monodomain model, initially in 1D and 2D domains \citep{herrero2022eppinn} and subsequently on 3D biventricular geometries \citep{conf/MICCAI/ye2023}. However, conventional PINNs typically require case-specific optimization, which limits their computational efficiency and generalizability to unseen patients. Recent advances have addressed this limitation through meta-learning, enabling rapid personalization from sparse measurements \citep{conf/MICCAI/toloubidokhti2025}, and physics-informed neural operators, enabling generalization across propagation scenarios and mesh resolutions \citep{lydon2025}. Nevertheless, these approaches impose physical constraints through soft loss regularization, which does not guarantee physical consistency and can be challenging to optimize for nonlinear reaction-diffusion systems \citep{journal/ANIPS/krishnapriyan2021}. Moreover, their evaluation remains largely restricted to simplified isotropic tissue models, neglecting the interplay between myocardial fibre anisotropy, anatomical variability, and scar distribution that governs cardiac EP \citep{journal/FiP/minchole2019,journal/doste2019,journal/NC/arevalo2016}.

An alternative direction is hybrid physics-neural modelling, which integrates mechanistic components with learnable operators to capture dynamics beyond prescribed physical models \citep{book/kashtanova2022}. HyPer-EP \citep{arxiv/jiang2024} combined mechanistic and neural components with meta-learning for patient-specific adaptation, while HAPI-EP \citep{conf/vadhavkar2026hapi} extended this paradigm to biventricular geometries by coupling Aliev-Panfilov reaction kinetics with attention-based graph diffusion. Although these approaches provide a more explicit representation of the underlying physiology, their mechanistic components remain largely prescribed, while the learned operators do not necessarily preserve essential physical properties by construction. Furthermore, their reliance on patient-specific measurements for model adaptation limits their applicability to forward simulation on unseen anatomies. 

\subsection{Graph-Based Learning for Physical Dynamics on Meshes} \label{sec:related_graph_dynamics}

Graph neural networks (GNNs) provide a flexible framework for learning physical dynamics on unstructured meshes, making them particularly suitable for patient-specific cardiac geometries reconstructed from clinical imaging \citep{journal/ABME/crozier2016,journal/nature/arevalo2016}. Through local message passing and shared network parameters, GNNs facilitate generalization across geometries and mesh resolutions \citep{conf/ICML/cao2023,conf/pfaff2020}. MeshGraphNets \citep{conf/pfaff2020} demonstrated this capability in structural and fluid mechanics, with subsequent multi-scale and attention-based architectures further improving computational scalability \citep{conf/fortunato2022,conf/nabian2024,conf/iparraguirre2026}. To improve physical consistency beyond purely data-driven learning, physics-informed GNNs incorporate discretized PDE residuals into the training objective \citep{journal/CMAM/wurth2024,book/de2024,journal/ASC/xiang2024}. More recently, physics-encoded architectures have integrated physical principles directly into network operations, including PDE operator decomposition \citep{conf/AAAI/yang2026,journal/JCP/lan2023dosnet,conf/ICML/karlbauer2022,conf/ICLR/zeng2025}, diffusion-based message passing, and structure- or conservation-preserving dynamics \citep{conf/iclm/chamberlain2021,journal/ANIPS/gruber2023,conf/ICML/hansen2023}. In particular, PhyMPGN \citep{conf/ICLR/zeng2025} employed a learnable Laplace-Beltrami operator for diffusion and a GNN for the remaining PDE terms. However, these architectures have primarily been developed for general physical systems, without explicitly addressing the nonlinear cellular excitation and recovery dynamics characteristic of cardiac EP.

In cardiac EP, graph-based neural surrogates have recently been explored for efficient simulation. AGATA \citep{conf/ICFIM/morier2025lagata} employed graph attention to learn electrical dynamics directly from data, while the graph-based surrogate in the GraphEP benchmark \citep{arxiv/baldan2026} incorporated prescribed Aliev-Panfilov reaction kinetics for left-ventricular EP simulation, without explicitly enforcing the physical properties of the learned diffusion operator. Our preliminary study, PI-GNN \citep{conf/MICCAI/tran2026}, incorporated monodomain residuals into the training objective of a recurrent MeshGraphNet, demonstrating generalization across unseen biventricular geometries and mesh resolutions. However, its reliance on loss-based physical constraints leaves the predicted dynamics susceptible to error accumulation during long-horizon simulation, particularly during repolarization. 
% These limitations motivate the development of graph-based neural surrogates that explicitly model cellular reaction and spatial diffusion, preserve their essential physical properties by architectural design, and enable accurate and stable long-horizon EP simulation across patient-specific anatomies and mesh resolutions.

% =========================================================
\section{Method}\label{sec:method}
%=========================================================

\begin{figure}[t]
    \centering
    \includegraphics[width=\columnwidth]{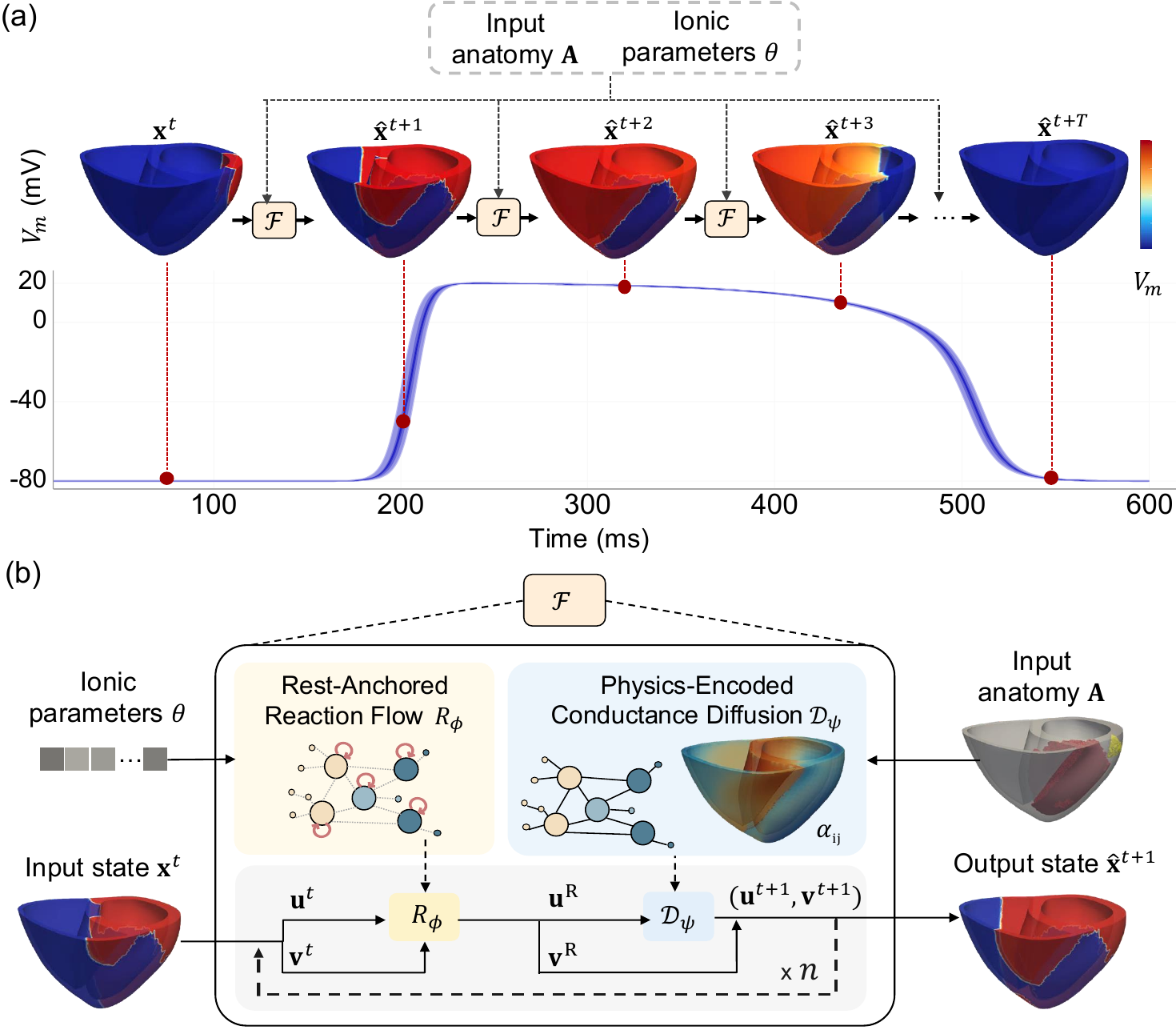}
    \caption{Overview of PhysCARD.
(a)~The neural simulator $\mathcal{F}$ predicts the cardiac state autoregressively, one frame at a time, from the post-stimulus state, conditioned on the anatomy $\mathbf{A}$ (geometry, fibres, conductivities and scar) and the ionic parameters $\boldsymbol{\theta}$ of the cell model; the trace shows transmembrane potential $V_m$ at one node as an example.
(b)~One step of $\mathcal{F}$: the diffusion module $\mathcal{D}_\psi$, informed by $\mathbf{A}$, and the reaction module $R_\phi$, informed by $\boldsymbol{\theta}$, alternate over $n$ substeps.}
    \label{fig:overview}
\end{figure}

\subsection{Problem Formulation}\label{sec:method:problem_formulation}

Given a patient-specific cardiac mesh and its post-stimulus EP state, our goal is to predict the electrical dynamics from activation through repolarization. Let $\mathcal{G}=(\mathcal{V},\mathcal{E})$ denote the cardiac mesh with $N$ nodes, and $\mathbf{A}$ be its anatomical descriptors, including geometry, fibre and sheet orientations, anisotropic conductivities, and scar distribution. The nodes comprise viable myocardium $\mathcal{V}_{\mathrm{myo}}$ and electrically insulating, non-excitable scar $\mathcal{V}_{\mathrm{scar}}$.
The EP dynamics of viable tissue follow the monodomain equation coupled with the Aliev-Panfilov model~\citep{journal/aliev1996}:
\begin{align}
    \frac{\partial u}{\partial t}
    &= \nabla\cdot\left(\mathbf{D}\nabla u\right)
    +\beta\left[k\,u(u-a)(1-u)-uv\right],
    \label{eq:ap_u}\\
    \frac{\partial v}{\partial t}
    &= \frac{\epsilon(u,v)}{\beta}
    \left[-v-k\,u(u-a-1)\right].
    \label{eq:ap_v}
\end{align}
Here, $V_m$ is the transmembrane potential, and $u\in[0,1]$ is its normalized form; $v$ is the recovery variable, and $\mathbf{D}$ is the anatomy-dependent anisotropic conductivity tensor. The parameters $k$ and $a$ govern excitation kinetics and the excitability threshold, while $\epsilon(u,v)=\epsilon_0+\mu_1v/(\mu_2+u)$ controls recovery kinetics. The parameter $\beta$ scales the excitation and recovery dynamics. The ionic parameters $\boldsymbol{\theta}=(k,a,\epsilon_0,\mu_1,\mu_2,\beta)$ are shared within each mesh but may vary across patients.
Defining the nodal state as $\mathbf{x}_i^t=[u_i^t,v_i^t]^\top$ and the whole-mesh state as $\mathbf{x}^t=[\mathbf{u}^t,\mathbf{v}^t]\in\mathbb{R}^{N\times2}$, we aim to learn a neural simulator $\mathcal{F}$ that advances the EP state over a time interval $\Delta t$:
\begin{equation}
    \hat{\mathbf{x}}^{t+1}
    =
    \mathcal{F}\big(
    \hat{\mathbf{x}}^t;
    \mathcal{G},\mathbf{A},\boldsymbol{\theta}
    \big),
    \label{eq:rollout}
\end{equation}
which is applied autoregressively from the post-stimulus state $\mathbf{x}^{t_0}$ over $(T-t_0)/\Delta t$ frames to predict the complete EP trajectory up to the final time $T$, as presented in Fig.~\ref{fig:overview}~(a).

\begin{figure*}[t]
    \centering
    \includegraphics[width=0.82\textwidth]{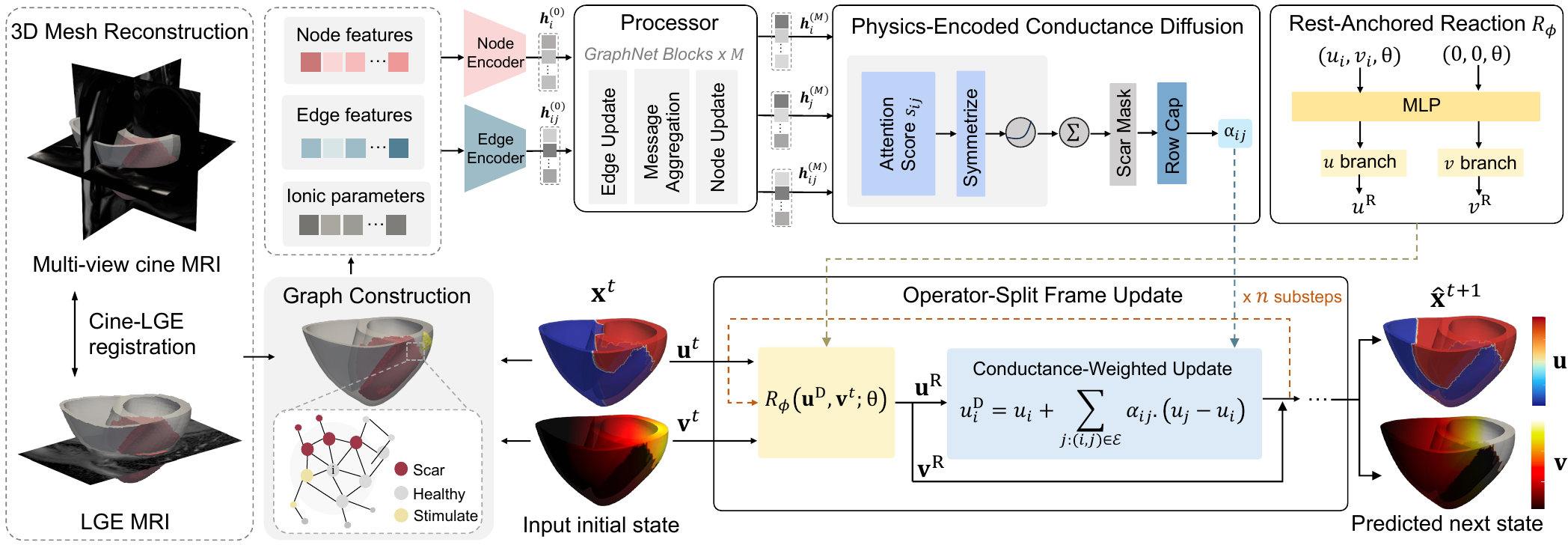}
    \caption{Network architecture of PhysCARD for one-step electrophysiology prediction. LGE: late gadolinium enhancement; MLP: multilayer perceptron.}
    \label{fig:architecture}
\end{figure*}

Rather than learning the state transition directly, PhysCARD follows the reaction-diffusion decomposition of the governing equations, as illustrated in Fig.~\ref{fig:overview} (b). Specifically, the rest-anchored reaction module $\mathcal{R}_\phi$ learns local excitation and recovery dynamics conditioned on the ionic parameters $\boldsymbol{\theta}$ (Sec.~\ref{sec:method:reaction}), while the physics-encoded diffusion module $\mathcal{D}_\psi$ learns anatomy-dependent electrical coupling through constrained graph attention (Sec.~\ref{sec:method:diffusion}). As illustrated in Fig.~\ref{fig:architecture}, each simulation step alternates the two modules over $n$ substeps: $\mathcal{R}_\phi$ first updates the transmembrane potential and recovery variable at each node, followed by $\mathcal{D}_\psi$, which propagates the updated potential across the cardiac mesh while preserving the recovery variable. The substep division here lets the wavefront advance at the physiological conduction velocity and keeps reaction and diffusion tightly coupled at the wavefront, which cannot be guaranteed by a single direct 1-ms prediction step. Unlike hybrid approaches combining mechanistic reaction models with learned diffusion~\citep{conf/vadhavkar2026hapi}, PhysCARD learns both operators and integrates them through physics-grounded operator splitting, with a staged learning strategy detailed in Sec.~\ref{sec:method:learning}.

%---------------------------------------------------------
\subsection{Rest-Anchored Reaction Flow}\label{sec:method:reaction}
%---------------------------------------------------------

\begin{figure}[t]
    \centering
    \includegraphics[width=0.98\columnwidth]{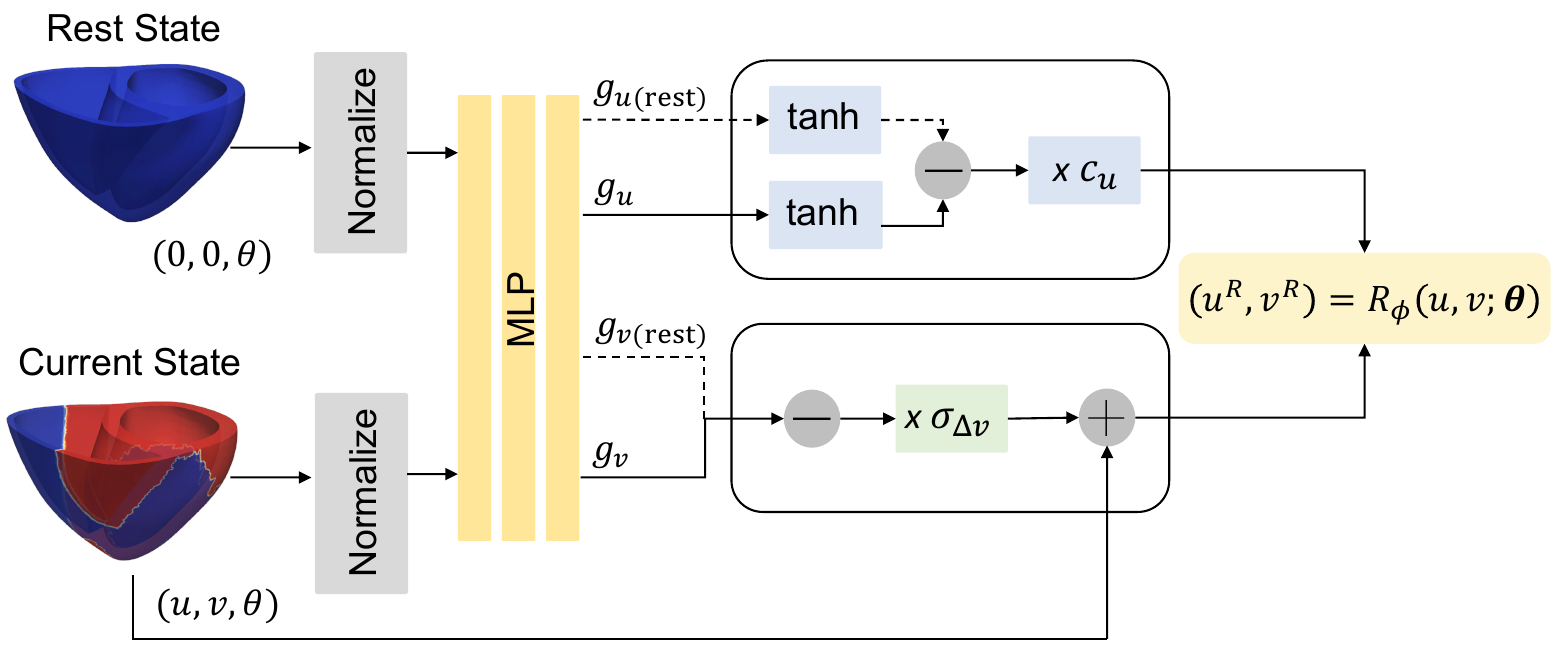}
    \caption{Rest-anchored reaction flow module $R_\phi$.}
    \label{fig:reaction}
\end{figure}

To model local excitation and recovery dynamics, we introduce a reaction flow module $\mathcal{R}_\phi$, shared across all viable myocardial nodes and conditioned on the ionic parameters $\boldsymbol{\theta}$. Let $\Phi_R(u,v;\boldsymbol{\theta})$ denote the reaction flow map obtained by integrating Eqs.~\eqref{eq:ap_u}-\eqref{eq:ap_v} without diffusion over a substep $\tau$. Following flow-map learning~\citep{journal/JCP/qin2019}, $\mathcal{R}_\phi$ directly approximates this state transition rather than predicting instantaneous reaction rates. This avoids explicit integration of the rapidly changing reaction rates during the action-potential upstroke, enabling efficient state updates in a single network evaluation.

As illustrated in Fig.~\ref{fig:reaction}, $\mathcal{R}_\phi$ is implemented as a multilayer perceptron (MLP) with two outputs $(g_u,g_v)$, evaluated at both the current state and the resting state:
\begin{equation}
    \begin{aligned}
        u^{R} &= c_u\big[\tanh g_u(u,v;\boldsymbol{\theta}) - \tanh g_u(0,0;\boldsymbol{\theta})\big],\\
        v^{R} &= v + \sigma_{\Delta v}\big[g_v(u,v;\boldsymbol{\theta}) - g_v(0,0;\boldsymbol{\theta})\big],
    \end{aligned}
    \label{eq:reaction}
\end{equation}
where $(u^{R},v^{R})=\mathcal{R}_\phi(u,v;\boldsymbol{\theta})$, $c_u$ is a fixed output scale, and $\sigma_{\Delta v}$ denotes the standard deviation of the one-frame recovery-variable increment. The two variables are parameterized differently to accommodate their distinct temporal dynamics: the rapidly changing potential $u$ is predicted directly with a bounded output, whereas the slowly evolving recovery variable $v$ is updated residually with increment normalization. This design captures rapid depolarization while resolving the gradual changes that govern repolarization.
Crucially, subtracting the network response at rest enforces $\mathcal{R}_\phi(0,0;\boldsymbol{\theta})=(0,0)$ by construction, ensuring that the resting state remains an exact equilibrium regardless of the learned parameters. This prevents spurious excitation in resting tissue and limits error accumulation during long-horizon autoregressive simulation. 
\subsection{Physics-Encoded Conductance Diffusion}\label{sec:method:diffusion}
%---------------------------------------------------------

To model electrical propagation across patient-specific cardiac meshes, we introduce a physics-encoded diffusion module $\mathcal{D}_\psi$ that learns the conductance between neighbouring nodes while preserving the physical properties of diffusion. Unlike conventional graph networks that directly predict potential changes, our module estimates edge-wise electrical coupling through an attention-based conductance head, followed by a fixed, physically constrained diffusion update, as presented in Fig.~\ref{fig:architecture}.
Following the anatomy-aware graph representation of our preliminary study~\citep{conf/MICCAI/tran2026}, node and edge features incorporate local potentials, scar labels, fibre orientations, anisotropic conductivities, and geometric connectivity. These features are processed by the encoder and $M$ residual message-passing blocks of MeshGraphNet~\citep{conf/pfaff2020}, producing node embeddings $\mathbf{h}^{(M)}_i$ and edge embeddings $\mathbf{h}^{M}_{ij}$. Rather than directly decoding potential increments, we introduce a conductance head that maps these embeddings to physically admissible edge conductances.

Specifically, following attention-based diffusion learning~\citep{conf/iclm/chamberlain2021,conf/vadhavkar2026hapi}, we compute an attention score vector for each edge:
\begin{equation}
    \mathbf{s}_{ij}
    =
    \mathrm{Attn}_{\psi}
    \left(
    \mathbf{h}^{(M)}_i,\mathbf{h}^{(M)}_j,\mathbf{h}^{(M)}_{ij}
    \right),
    \label{eq:score}
\end{equation}
where $\mathrm{Attn}_{\psi}$ denotes a learnable multi-head attention function based on scaled dot-product attention~\citep{journal/vaswani2017}. The resulting scores capture anatomy- and state-dependent interactions but do not inherently satisfy the physical constraints of diffusion.
To enforce physically consistent conduction, we impose four complementary constraints: \textit{symmetry} for balanced flux exchange, \textit{non-negativity} for gradient-driven diffusion, \textit{scar masking} for electrical insulation, and \textit{bounded conductance} for numerical stability. 
We first symmetrize the attention scores as
$\bar{\mathbf{s}}_{ij}=(\mathbf{s}_{ij}+\mathbf{s}_{ji})/2$
and map them to non-negative, scar-aware conductances:
\begin{equation}
    \tilde{\alpha}_{ij}
    =
    \mathds{1}\big[i,j\in\mathcal{V}_{\mathrm{myo}}\big]
    \mathbf{1}^{\top}
    \mathrm{softplus}
    \left(\bar{\mathbf{s}}_{ij}+\mathbf{b}\right),
    \label{eq:raw_conductance}
\end{equation}
where $\mathbf{b}$ denotes learnable attention biases, and $\mathbf{1}^{\top}$ sums the contributions across attention heads.
This formulation ensures symmetric, non-negative conductances while preventing electrical exchange across scar boundaries.
We further bound the total conductance of each node while preserving symmetry:
\begin{equation}
    \alpha_{ij}
    =
    \frac{\tilde{\alpha}_{ij}}
    {\max(1,S_i,S_j)},
    \qquad
    S_i=\sum_j\tilde{\alpha}_{ij},
    \label{eq:conductance}
\end{equation}
ensuring $\sum_j\alpha_{ij}\leq 1$. Together with non-negativity, this constraint establishes a discrete maximum principle, preventing the diffusion update from exceeding the local potential range.
The predicted conductances are then applied through a fixed diffusion operator:
\begin{equation}
    u_i^\ast
    =
    u_i+
    \sum_{j:(i,j)\in\mathcal{E}}
    \alpha_{ij}(u_j-u_i),
    \label{eq:diffusion_update}
\end{equation}
where $\alpha_{ij}(u_j-u_i)$ represents the potential exchange between neighbouring nodes. Importantly, these physical properties hold by construction, independently of the learned parameters or input states, including those encountered during long autoregressive rollouts. The difference-based formulation also preserves uniform potentials, complementing the resting-state constraint of $\mathcal{R}_\phi$ (Sec.~\ref{sec:method:reaction}).

\subsection{Physics-Grounded Operator-Split Learning}\label{sec:method:learning}
%---------------------------------------------------------

The reaction module $\mathcal{R}_\phi$ models local cellular dynamics, while the diffusion module $\mathcal{D}_\psi$ governs spatial electrical propagation. Each simulation frame is advanced through $n$ substeps of duration $\tau=\Delta t/n$, applying reaction followed by diffusion (Fig.~\ref{fig:architecture}). Edge conductances are computed once per frame and reused across substeps for efficient inference. 
To prevent the two modules from compensating for errors of each other and improve long-horizon stability, we adopt a three-stage physics-grounded learning strategy that progressively learns cellular kinetics, spatial conduction, and their coupled dynamics.

\subsubsection{Stage A: Learning Cellular Reaction Dynamics}

We first train $\mathcal{R}_\phi$ independently of the cardiac mesh, exploiting the fact that cellular reaction dynamics can be learned directly from the Aliev-Panfilov model. Let $\Phi_R(\mathbf{x};\boldsymbol{\theta})$ denote the reference reaction flow over one substep, obtained through fine-step numerical integration. We train $\mathcal{R}_\phi$ using
\begin{equation}
\begin{aligned}
\mathcal{L}_{\mathrm{cell}}
=&\;
\mathbb{E}\left[
\|\mathcal{R}_\phi(\mathbf{x})-\Phi_R(\mathbf{x})\|_\sigma^2
\right] \\
&+\lambda_J\mathbb{E}\left[
\|\partial_{\mathbf{x}}\mathcal{R}_\phi
-\partial_{\mathbf{x}}\Phi_R\|_F^2
\right]
+\mathcal{L}_{\mathrm{roll}},
\end{aligned}
\label{eq:cell_loss}
\end{equation}
where dependence on $\boldsymbol{\theta}$ is omitted for brevity, and $\|\cdot\|_\sigma$ denotes component-wise normalization by the standard deviations of state increments. The three terms constrain the reaction flow, its sensitivity to initial states, and its accumulated error over repeated predictions. In particular, Jacobian matching improves sensitivity near the excitation threshold, while $\mathcal{L}_{\mathrm{roll}}$ encourages accurate recovery dynamics over a complete action-potential cycle. Training states are sampled from physiological ranges, simulated trajectories, and regions near the excitation threshold to capture diverse cellular dynamics.

\subsubsection{Stage B: Learning Physics-Consistent Conduction}

With $\mathcal{R}_\phi$ frozen, we learn $\mathcal{D}_\psi$ within the coupled reaction-diffusion update. This prevents the reaction module from compensating for inaccurate spatial conduction. To improve robustness beyond observed trajectories, we perturb the input state as $\tilde{\mathbf{x}}^t=\mathbf{x}^t+\boldsymbol{\xi}$ and introduce a physics teacher $\Phi$,  which computes the physically consistent next state from the states generated by the model itself.
Specifically, $\Phi$ uses the governing Aliev-Panfilov reaction equations and finite-element
diffusion, with explicit Euler integration for reaction and Crank-Nicolson integration for diffusion. Starting from a ground-truth state $\mathbf{x}^t$, the model is first rolled out for $K$ steps without gradient, $\hat{\mathbf{x}}^{t+K}=\mathcal{F}^{K}(\mathbf{x}^t)$, and $\Phi$ then advances this predicted state by one frame. Unlike direct supervision using the clean next state, which is available only along the reference trajectories, the teacher provides the correct physical evolution of the states that the model visits during autoregressive rollout, reducing the accumulation of errors.

Since $\mathcal{D}_\psi$ updates only the potential, the diffusion module is optimized using a state prediction loss on $u$:
\begin{equation}
\mathcal{L}_{\mathrm{data}}
=
\frac{1}{|\mathcal{V}_{\mathrm{myo}}|}
\sum_{i\in\mathcal{V}_{\mathrm{myo}}}
\left(
\frac{
[\mathcal{F}(\tilde{\mathbf{x}}^t)]_{u,i}
-
u_i^{t+1}
}{\sigma_{\Delta u}}
\right)^2,
\label{eq:data_loss}
\end{equation}
where $\sigma_{\Delta u}$ is the standard deviation of potential increments. Additionally, we constrain the predicted potential using the physics teacher:
\begin{equation}
\mathcal{L}_{\mathrm{phys}}
=
\frac{1}{|\mathcal{V}_{\mathrm{myo}}|}
\sum_{i\in\mathcal{V}_{\mathrm{myo}}}
\left(
\frac{
[\mathcal{F}(\hat{\mathbf{x}}^{t+K})]_{u,i}
-
[\Phi(\hat{\mathbf{x}}^{t+K})]_{u,i}
}{\sigma_{\Delta u}}
\right)^2,
\label{eq:physics_loss}
\end{equation}
This loss explicitly constrains conduction on model-generated states, while the physics teacher is used only during training and introduces no computational overhead at inference. The Stage B objective is $\mathcal{L}_{B}=\mathcal{L}_{\mathrm{data}}
+\lambda_P\mathcal{L}_{\mathrm{phys}}$.

\subsubsection{Stage C: Joint Reaction-Diffusion Adaptation}

Finally, both modules are jointly fine-tuned to account for their interaction during tissue-level propagation. Since the reaction module now receives states influenced by spatial diffusion, its input distribution differs from that encountered during isolated cellular training. However, unconstrained adaptation could distort the learned reaction kinetics to compensate for diffusion errors.
To preserve physiological consistency, we introduce a reaction distillation loss:
\begin{equation}
\mathcal{L}_{\mathrm{dist}}
=
\mathbb{E}\left[
\left\|
\mathcal{R}_\phi(\mathrm{sg}[\mathbf{x}_{\mathrm{in}}])
-
\Phi_R(\mathrm{sg}[\mathbf{x}_{\mathrm{in}}])
\right\|_\sigma^2
\right],
\label{eq:distillation_loss}
\end{equation}
where $\mathbf{x}_{\mathrm{in}}$ denotes the reaction inputs encountered during the coupled substeps, $\Phi_R$ is the reaction part of $\Phi$, $\|\cdot\|_\sigma$ is the error normalized by the standard deviation of state increments, and $\mathrm{sg}[\cdot]$ is the stop-gradient operator. This term allows the reaction module to adapt to tissue-level states while remaining consistent with the underlying ionic dynamics.
The final objective combines data fidelity, physics consistency, and reaction distillation:
\begin{equation}
\mathcal{L}_{C}
=
\mathcal{L}_{\mathrm{data}}
+
\lambda_P\mathcal{L}_{\mathrm{phys}}
+
\lambda_D\mathcal{L}_{\mathrm{dist}}.
\label{eq:joint_loss}
\end{equation}
Together, the three stages establish physiologically grounded reaction and diffusion operators before jointly adapting them for accurate long-horizon EP simulation.

\section{Experiments and Results}\label{sec:exp}
%=========================================================

%---------------------------------------------------------
\subsection{Dataset Construction and Simulation}\label{sec:exp:data}
%---------------------------------------------------------

We used paired short-axis late gadolinium enhancement (LGE) and multi-view cine MRI from 37 post-myocardial infarction patients acquired on a 3.0\,T Siemens scanner at the National University Hospital, Singapore. Biventricular surface meshes were reconstructed from cine MRI using biv-me~\citep{journal/MedIA/dillon2026,conf/mauger2018}. Myocardial scars were manually delineated on LGE MRI and mapped to the cine-derived meshes via registration~\citep{journal/arxiv/lyu2026}. The slice-wise annotations were reconstructed in 3D using inter-slice interpolation and anisotropic Gaussian diffusion, followed by transmural extension using consistent biventricular coordinates~\citep{journal/MedIA/schuler2021}. Volumetric meshes with rule-based fibre and sheet orientations and AHA segment labels were generated following~\citet{journal/doste2026}, and resampled using meshtool at nominal resolutions of 1000, 800, and 500\,\textmu m.

Reference EP trajectories were simulated using openCARP~\citep{journal/CMPB/plank2021} with the monodomain equation and Aliev-Panfilov model ($a=0.15$, $\epsilon_0=0.003$, $\mu_1=1.4$, $\mu_2=0.5$, $\beta=12.9$). Dense scar was treated as non-excitable and electrically insulating, without modelling a border zone. The solver employed a consistent mass matrix and operator splitting with Crank-Nicolson diffusion, using a 25\,\textmu s time step and 1\,ms output interval. Conductivities were calibrated separately at each resolution using tuneCV \citep{conf/CinC/costa2013} to achieve longitudinal and transverse conduction velocities of 0.60\,m/s and 0.39\,m/s, respectively. Each simulation was initiated by a 2\,ms stimulus applied to a 3\,mm-radius endocardial patch, with up to two randomly selected AHA segments paced per anatomy and resolution. The resulting trajectories contained transmembrane potential $V_m$ ($V_m=100u-80$) and recovery variable $v$, with scar nodes fixed at $V_m=-80$\,mV and $v=0$.
The dataset was partitioned at the patient level, ensuring no overlap across training, validation, and test sets. The 1000 and 800\,\textmu m meshes were used for model development and in-distribution evaluation, while the unseen 500\,\textmu m resolution was reserved for assessing cross-resolution generalization. 
% Normalization statistics were computed exclusively from the training data. 
Table~\ref{tab:data} summarizes the mesh characteristics, simulation settings, and data partitions.

\begin{table}[t]
    \centering
    \caption{Dataset summary per nominal mesh resolution. APD: action potential duration.}
    \label{tab:data}
    \footnotesize
    \setlength{\tabcolsep}{1pt}
    \begin{adjustbox}{max width=\textwidth}
    \begin{tabular}{lccc}
        \toprule
        & 1000\,\textmu m & 800\,\textmu m & 500\,\textmu m \\
        \midrule
        Mean edge length (\textmu m)      & $1204\pm8$  & $884\pm13$  & $553\pm1$ \\
        Nodes per mesh ($\times10^3$)     & $134\pm25$  & $319\pm54$  & $1396\pm232$ \\
        Aliev-Panfilov $k$               & 0.8 & 1.3 & 1.3 \\
        APD$_{90}$ (ms)                   & 427 & 295 & 296 \\
        Total activation time (ms)        & $193\pm30$  & $228\pm35$  & $177\pm19$ \\
        Simulated duration (ms)           & 650 & 510 & 600 \\
        Anatomies (train/ val/ test)        & 12 / 3 / 10	 & 12 / 3 / 10	 & 0 / 0 / 5 \\
        Trajectories (train/ val/ test)     & 24 / 6 / 10	 & 21 / 4 / 10 & 0 / 0 / 5 \\
        \bottomrule
    \end{tabular}
    \end{adjustbox}
\end{table}

\subsection{Ground Truth and Evaluation}

We evaluated PhysCARD against reference openCARP simulations in terms of state prediction accuracy, EP fidelity, and physical consistency, considering both one-step predictions from ground-truth states and autoregressive rollouts. State prediction accuracy was quantified using root mean squared error (RMSE) and Pearson correlation coefficient (PCC) for the transmembrane potential $V_m$ and recovery variable $v$, with the RMSE of $v$ additionally normalized by its range in the training data at each mesh resolution (NRMSE). EP fidelity was assessed using local activation time (LAT), defined as the first upward crossing of $-10$\,mV, repolarization time (RT), defined at 90\% repolarization, and action potential duration ($\mathrm{APD}=\mathrm{RT}-\mathrm{LAT}$), with mean absolute errors (MAEs) computed over nodes where each event occurred in both prediction and ground truth. Organ-level markers included the Pearson correlation $r$ of LAT and RT maps, the error in the time to 90\,\% activation (AT90), the mean signed APD error (APD bias), and the percentages of nodes with missed repolarization or spurious re-excitation. Node-level agreement was summarized by the Bland-Altman bias with 95\,\% limits of agreement (LoA) and concordance correlation coefficient (CCC). Physical consistency was quantified by the solver-consistency error $E_\Phi$, defined as the one-step RMSE between $\mathcal{F}(\hat{\mathbf{x}}^t)$ and the reference solver $\Phi(\hat{\mathbf{x}}^t)$ on autoregressively predicted states. The learned diffusion operator was further evaluated against finite-element diffusion using cosine similarity and wavefront-direction agreement.
% , together with assessments of charge balance, flux conservation, conductance symmetry and non-negativity, and violations of the discrete maximum principle. 
% To evaluate the significance, differences between PhysCARD and each baseline were assessed with two-sided paired Wilcoxon signed-rank tests over the test trajectories, with Holm correction across all baselines and metrics.

%---------------------------------------------------------
\subsection{Implementation Details}\label{sec:exp:implementation}
%---------------------------------------------------------

The framework was implemented in PyTorch with PyTorch Geometric and trained on a single NVIDIA 6000D GPU. The MeshGraphNet used two-layer ReLU perceptrons with layer normalization and a latent size of 64, followed by $M=5$ residual message-passing blocks with sum aggregation. The conductance head used two attention heads of dimension 16, with head biases initialized to $-3$. The reaction module was a three-layer perceptron with 128 SiLU units and output scale $c_u=1.5$, and each frame was advanced by $n=2$ substeps of $\tau=0.5$\,ms with reaction followed by diffusion.

% \paragraph{Training}
In Stage~A, $R_\phi$ was trained for 30 epochs of 400 steps with Adam, a learning rate decaying exponentially from $10^{-3}$ to $10^{-5}$, batches of 4096 states, and gradient clipping at norm 1.
The labels $\Phi_R$ were computed on the fly by 20 explicit Euler steps of 0.025\,ms, and states were drawn 50\% from ground-truth states perturbed by Gaussian noise, 30\% uniformly over the range of the training data, and 20\% near the threshold, with the ionic parameters of one of the two resolutions.
The Jacobian term used $\lambda_J=0.1$ on a quarter of each batch, and the rollout term compared 512 single-cell trajectories over horizons of 8, 32, 128 and 400\,ms, introduced at epochs 1, 6, 12 and 20.
Stages~B and~C ran for 30 and 20 epochs, respectively, with Adam, no weight decay, and a learning rate decaying exponentially from $10^{-4}$ to $10^{-6}$ over both stages, which was reduced tenfold for $R_\phi$ in Stage~C. The on-policy states for the physics loss were generated by rolling out the model for $K=4$ steps from the ground-truth state without gradient.
Each epoch drew 768 one-step training pairs from frames $t\geq3$\,ms, one graph per step, and perturbed the viable nodes with independent Gaussian noise of standard deviation 1.07\,mV on $V_m$ and $3.7\times10^{-4}$ on $v$.
The physics loss was evaluated on every second pair with $t\geq5$\,ms, and $\lambda_P=\lambda_D=1$.
The teacher solved the Crank-Nicolson system with 40 Jacobi-preconditioned conjugate-gradient iterations, and reproduced the next ground-truth frame with a mean RMSE of $2.6\times10^{-5}$\,mV over 90 training states.
Every five epochs, the model was evaluated by rollouts on the validation trajectories, and we retained the checkpoint with the lowest validation $V_m$ RMSE, averaged within and then across resolutions.
% Stages~B and~C took 4.9\,h and 3.4\,h, respectively, and all reported results use a single training seed. 

\begin{table*}[!t]
\centering
\caption{Quantitative comparison on unseen patient biventricular meshes within the in-distribution test set (20 anatomies at 1000 and 800\,\textmu m).
$V_m=100u-80$; 1-step: each frame is predicted from the ground-truth state of the preceding frame; Rollout: each frame is predicted from the model's own preceding prediction, starting from the ground truth at $t_0=10$ ms;
(N)RMSE: (normalized) root mean squared error; PCC: Pearson correlation coefficient; MAE: mean absolute error; LAT: local activation time; RT: repolarization time; --: not available as more than 97\% of the nodes did not repolarize.}
\label{tab:main}

\footnotesize
\setlength{\tabcolsep}{3pt}
\renewcommand{\arraystretch}{1.15}

\begin{adjustbox}{max width=\textwidth}
\begin{tabular}{l cccc cccc ccc}
\toprule

\multirow{3}{*}{Method}
& \multicolumn{4}{c}{Transmembrane potential $V_m$}
& \multicolumn{4}{c}{Recovery variable $v$}
& \multicolumn{3}{c}{Timing MAE (ms)~$\downarrow$} \\

\cmidrule(lr){2-5}
\cmidrule(lr){6-9}
\cmidrule(lr){10-12}

& \multicolumn{2}{c}{RMSE (mV)$~\downarrow$}
& \multicolumn{2}{c}{PCC$~\uparrow$}
& \multicolumn{2}{c}{NRMSE (\%)~$\downarrow$}
& \multicolumn{2}{c}{PCC~$\uparrow$}
& \multirow{2}{*}{LAT}
& \multirow{2}{*}{RT}
& \multirow{2}{*}{APD} \\

\cmidrule(lr){2-3}
\cmidrule(lr){4-5}
\cmidrule(lr){6-7}
\cmidrule(lr){8-9}

& 1-step & Rollout
& 1-step & Rollout
& 1-step & Rollout
& 1-step & Rollout
& & & \\

\midrule

GINO~\citep{journal/ANIPS/li2023gino}
& 5.55 $\pm$ 1.89 & 45.7 $\pm$ 11.8 & .911 $\pm$ .089 & $-$.018 $\pm$ .181 & 4.51 $\pm$ 1.78 & 24.7 $\pm$ 3.64 & .892 $\pm$ .142 & $-$.034 $\pm$ .224 & 40.5 $\pm$ 16.3 & -- & -- \\

AGATA~\citep{conf/ICFIM/morier2025lagata}
& 2.26 $\pm$ .919 & 39.2 $\pm$ 8.67 & .992 $\pm$ .006 & .158 $\pm$ .023 & 2.23 $\pm$ 1.21 & 22.6 $\pm$ 1.22 & .988 $\pm$ .003 & .065 $\pm$ .035 & 24.0 $\pm$ 10.5 & -- & -- \\

PhyMPGN~\citep{conf/ICLR/zeng2025}
& .575 $\pm$ .154 & 17.6 $\pm$ 4.32 & .998 $\pm$ .001 & .640 $\pm$ .113 & .059 $\pm$ .008 & 12.1 $\pm$ 3.08 & 1.00 $\pm$ .000 & .737 $\pm$ .075 & 1.38 $\pm$ 0.64 & 28.0 $\pm$ 12.7 & 28.4 $\pm$ 12.0 \\

MeshGraphNet~\citep{conf/pfaff2020}
& .413 $\pm$ .080 & 23.3 $\pm$ 8.91 & .998 $\pm$ .002 & .524 $\pm$ .223 & .032 $\pm$ .003 & 9.88 $\pm$ 2.04 & 1.00 $\pm$ .000 & .564 $\pm$ .243 & 19.9 $\pm$ 12.5 & 27.9 $\pm$ 9.88 & 25.5 $\pm$ 15.0 \\
% PI-GNN w/o physics loss
% & .402 $\pm$ .063
% & 20.9 $\pm$ 7.92
% & .998 $\pm$ .001
% & .554 $\pm$ .245
% & .019 $\pm$ .002
% & 9.73 $\pm$ 4.60
% & 1.00 $\pm$ .000
% & .578 $\pm$ .307
% & 15.7 $\pm$ 13.1
% & 30.6 $\pm$ 23.0
% & 19.3 $\pm$ 12.8 \\
PI-GNN~\citep{conf/MICCAI/tran2026}
& .371 $\pm$ .077 & 16.1 $\pm$ 4.13 & .998 $\pm$ .001 & .720 $\pm$ .140 & .019 $\pm$ .001 & 8.89 $\pm$ 1.86 & 1.00 $\pm$ .000 & .663 $\pm$ .149 & 8.23 $\pm$ 10.7 & 24.3 $\pm$ 8.23 & 23.8 $\pm$ 8.67 \\

% \midrule
\textbf{PhysCARD (Ours)}
& \textbf{.338 $\pm$ .059} & \textbf{4.52 $\pm$ 2.21} & \textbf{1.00 $\pm$ .001} & \textbf{.955 $\pm$ .034} & \textbf{.015 $\pm$ .004} & \textbf{1.73 $\pm$ 1.14} & \textbf{1.00 $\pm$ .000} & \textbf{.990 $\pm$ .009} & \textbf{1.23 $\pm$ 0.72} & \textbf{3.71 $\pm$ 2.64} & \textbf{3.36 $\pm$ 3.04} \\
\bottomrule
\end{tabular}
\end{adjustbox}

\end{table*}

%---------------------------------------------------------
\subsection{Comparison Study} \label{sec:exp:comparison}
%---------------------------------------------------------

We compared PhysCARD with GINO~\citep{journal/ANIPS/li2023gino}, AGATA~\citep{conf/ICFIM/morier2025lagata}, PhyMPGN~\citep{conf/ICLR/zeng2025}, MeshGraphNet~\citep{conf/pfaff2020}, and PI-GNN~\citep{conf/MICCAI/tran2026}, our preliminary conference method. PI-GNN is a recurrent MeshGraphNet trained on multi-step rollouts with a physics loss. We adapted GINO and PhyMPGN from their public implementations, followed the published MeshGraphNet architecture, and reimplemented AGATA from its published description. GINO used its original encoder-Fourier neural operator-decoder architecture to predict the next $(V_m,v)$ state from the current state and conditioning inputs, with anatomy represented by a mesh signed-distance field. AGATA predicted $(V_m,v)$ from a 25-ms history of both variables, using the same conditioning inputs as MeshGraphNet. PhyMPGN used a Crank-Nicolson diffusion step on the anisotropic three-dimensional mesh; to keep its long rollouts stable, we standardized its inputs and outputs and trained it with a rollout curriculum and noise on both state variables. Each baseline followed its own training protocol, with checkpoints selected on the validation set. For rollout evaluation, all methods were initialized using ground-truth data available through $t_0=10$\, ms (25\,ms for AGATA, the length of its input history) and predicted autoregressively to the final frame. To compare all methods under the conditions they were trained for, Table~\ref{tab:main} reports the 20 in-distribution test trajectories at 1000 and 800\,\textmu m. The five 500\,\textmu m meshes, whose resolution was unseen in training and on which several baselines became unstable or failed to repolarize, are analyzed separately in Sec.~\ref{sec:exp:scalability}.

Table~\ref{tab:main} compares the six methods in one-step prediction and full-rollout simulation. PhysCARD, PI-GNN, MeshGraphNet, and PhyMPGN had similar one-step $V_m$ RMSEs (0.34-0.58\,mV), whereas their rollout RMSEs were 4.52, 16.1, 23.3, and 17.6\,mV, respectively. GINO and AGATA had larger errors at both horizons, with one-step RMSEs of 2.26-5.55\,mV and rollout RMSEs of at least 39.2\,mV. PhysCARD achieved a rollout $V_m$ Pearson correlation coefficient of 0.955 and reduced rollout $V_m$ RMSE by approximately 72\% relative to PI-GNN, while lowering the NRMSE of the recovery variable $v$ from 8.89\% to 1.73\%. For both $V_m$ and $v$, PhysCARD had significantly lower rollout errors than each baseline (paired Wilcoxon signed-rank tests, all Holm-adjusted $p<0.001$). Relative to PI-GNN, PhysCARD also reduced the MAEs in LAT, RT, and APD from 8.23, 24.3, and 23.8\,ms to 1.23, 3.71, and 3.36\,ms, respectively. The RT and APD errors were also significantly lower with PhysCARD than with PI-GNN, MeshGraphNet, or PhyMPGN ($p<0.001$). With its Crank-Nicolson diffusion step, PhyMPGN achieved a LAT error of 1.38\,ms, close to the 1.23\,ms observed with PhysCARD ($p=0.28$), while its errors in RT and APD were both approximately 28\,ms.

\begin{figure}[t]
    \centering
    \includegraphics[width=0.9\columnwidth]{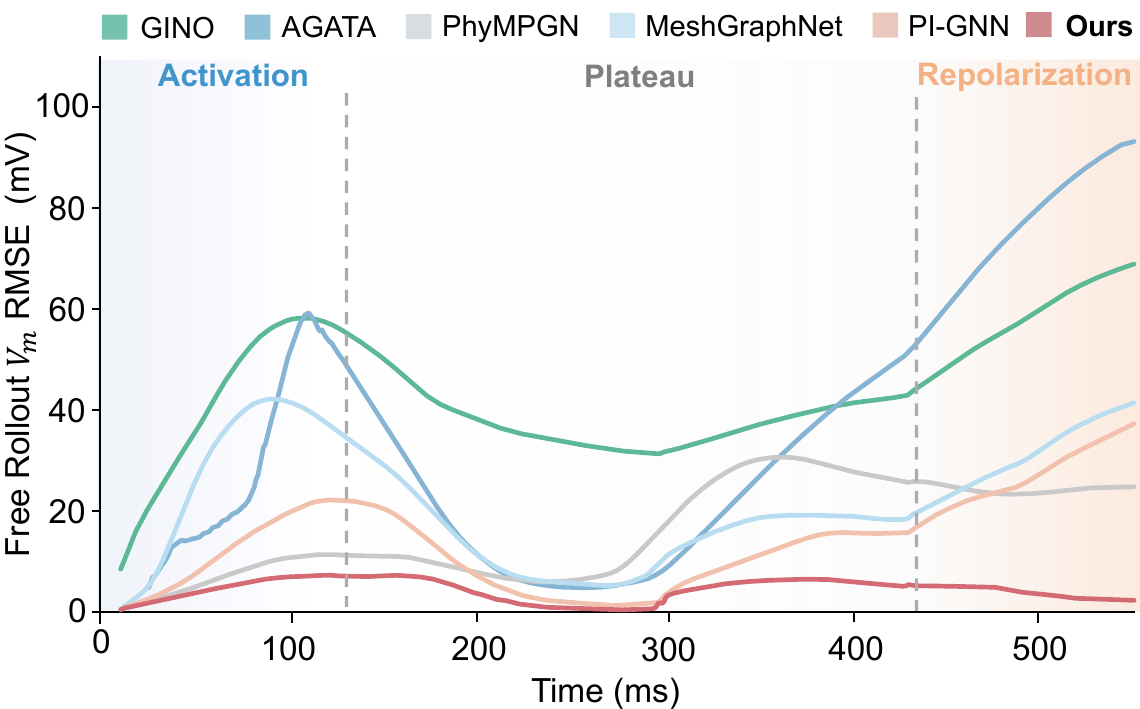}
    \vspace{-5mm}
    \caption{Temporal evolution of transmembrane potential (\(V_m\)) errors during free rollout on test trajectories. }
    \label{fig:vm_error}
\end{figure}
        
\begin{figure*}[!t]
    \centering
    \includegraphics[width=0.86\textwidth]{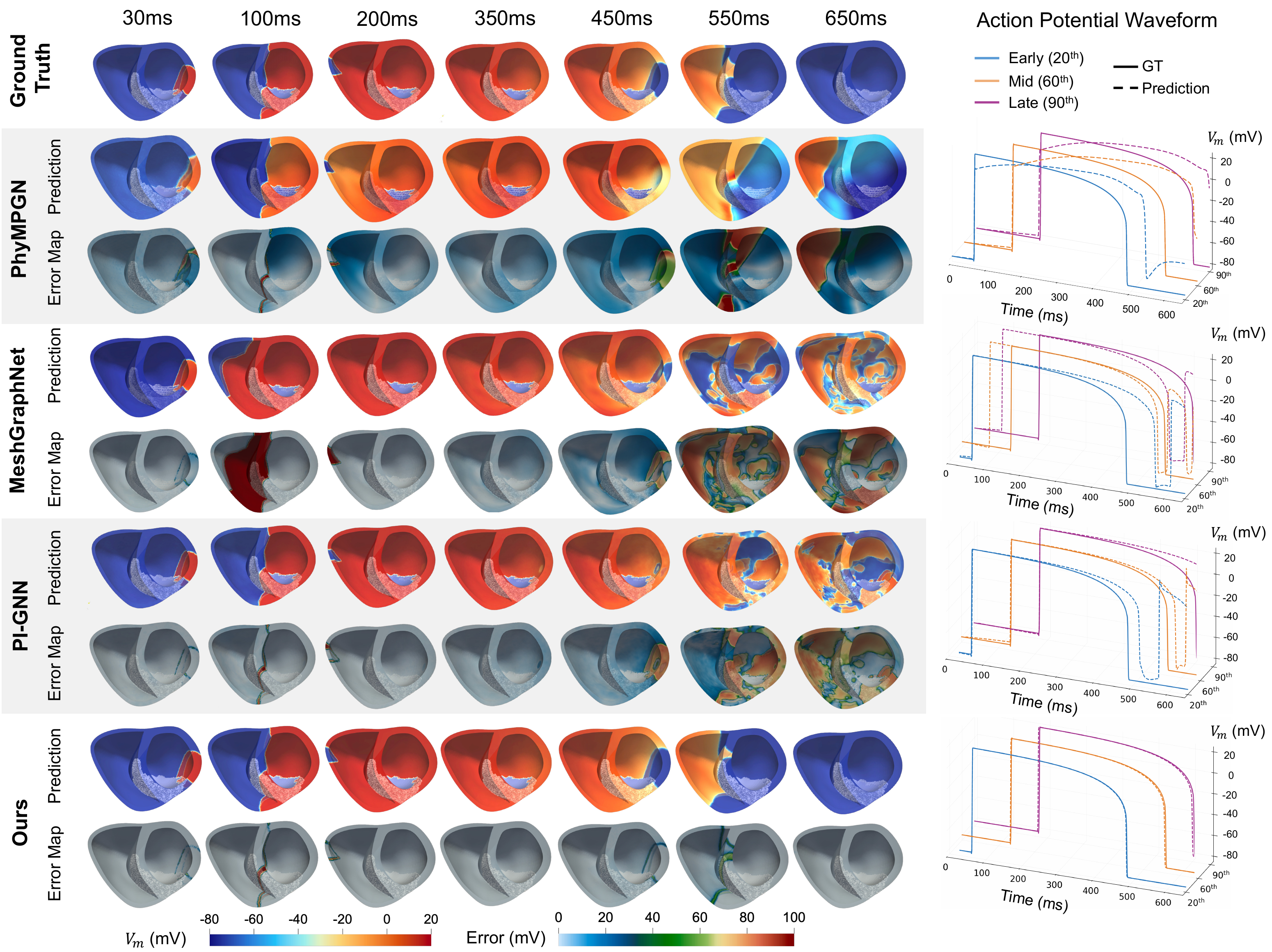}
    \caption{Qualitative comparison of cardiac electrophysiology simulation of different methods.
    Left: predicted transmembrane potentials (\(V_m\)) and absolute error maps at different times. 
    Right: action potential waveforms at nodes representing the 20th, 60th, and 90th percentiles of ground-truth (GT) activation times.}
    \label{fig:qualitative}
\end{figure*}

The temporal error profiles in Fig.~\ref{fig:vm_error} show when these differences emerged. PhysCARD maintained the lowest $V_m$ RMSE during activation, and the gap between PhysCARD and PI-GNN narrowed during the plateau.  As repolarization progressed, the errors of most baselines increased, while the PhysCARD error stayed below 7\,mV and decreased to about 2\,mV by the end of the rollout. In the representative trajectory shown in Fig.~\ref{fig:qualitative}, the spatial errors of PhysCARD were concentrated near the activation and repolarization fronts, while MeshGraphNet and PI-GNN produced patchy repolarization. At the illustrated nodes, PI-GNN and PhysCARD reproduced the upstroke similarly, but PI-GNN repolarized later and showed spurious re-excitation, whereas PhysCARD more closely followed the simulated return to rest. Together with the lower rollout error in $v$, these findings are consistent with more accurate recovery dynamics in PhysCARD.

\subsection{Ablation Study}\label{sec:exp:ablation}
%---------------------------------------------------------

\begin{table*}[t]
\centering
\caption{Ablation of PhysCARD components on the full test set (25 trajectories at three resolutions).
Repol.\ $V_m$: RMSE within $[\mathrm{RT}-30,\mathrm{RT}+15)$\,ms of each node; Missed repol.: repolarizing in the GT but not in the prediction; Re-excitation: excited again after recovery.}
\label{tab:ablation}
\footnotesize
\setlength{\tabcolsep}{3.5pt}
\renewcommand{\arraystretch}{1.15}
\begin{adjustbox}{max width=\textwidth}
\begin{tabular}{l cccc ccc ccc}
% \toprule
% & \multicolumn{4}{c}{State}
% & \multicolumn{3}{c}{Timing MAE (ms) $\downarrow$}
% & \multicolumn{3}{c}{Physical consistency (\%) $\downarrow$} \\
% \cmidrule(lr){2-5}\cmidrule(lr){6-8}\cmidrule(lr){9-11}
% Variant
% & $V_m$ RMSE (mV) $\downarrow$ & $V_m$ PCC $\uparrow$ & $v$ NRMSE (\%) $\downarrow$ & Repol.\ $V_m$ RMSE (mV) $\downarrow$
% & LAT & RT & APD
% & Re-excitation & Missed repol. & APD err.\ $>$10\,ms \\
% \midrule
\toprule
\multirow{3}{*}{Variant}
& \multicolumn{4}{c}{State}
& \multicolumn{3}{c}{Timing MAE (ms) $\downarrow$}
& \multicolumn{3}{c}{Physical consistency (\%) $\downarrow$} \\
\cmidrule(lr){2-5}\cmidrule(lr){6-8}\cmidrule(lr){9-11}
& $V_m$ RMSE (mV) $\downarrow$
& $V_m$ PCC $\uparrow$
& $v$ NRMSE (\%) $\downarrow$
& Repol.\ $V_m$ (mV) $\downarrow$
& LAT & RT & APD
& Re-excitation & Missed repol. & APD err.\ $>$10\,ms \\
\midrule
\multicolumn{11}{l}{\emph{Diffusion module}} \\
\quad w/o conductance head
& 5.46 $\pm$ 3.12
& 0.93 $\pm$ 0.06
& 2.24 $\pm$ 1.37
& 19.6 $\pm$ 10.2
& 1.59 $\pm$ 1.63
& 5.57 $\pm$ 3.52
& 4.13 $\pm$ 2.51
& \textbf{0.00 $\pm$ 0.00}
& $<$ 0.01
& \textbf{0.00 $\pm$ 0.00} \\
\quad w/o constrained attention
& 5.60 $\pm$ 2.55
& 0.93 $\pm$ 0.05
& 2.22 $\pm$ 1.23
& 19.3 $\pm$ 8.82
& 1.65 $\pm$ 1.31
& 5.24 $\pm$ 3.01
& \textbf{4.04 $\pm$ 2.29}
& $<$ 0.01
& $<$ 0.01
& 0.28 $\pm$ 0.74 \\
\midrule

\multicolumn{11}{l}{\emph{Reaction module}} \\
\quad w/o rest anchor
& 14.5 $\pm$ 11.6
& 0.79 $\pm$ 0.21
& 6.22 $\pm$ 2.56
& 39.8 $\pm$ 8.23
& 14.6 $\pm$ 25.8
& 20.78 $\pm$ 13.7
& 15.90 $\pm$ 5.21
& $<$ 0.01
& 1.63 $\pm$ 4.69
& 69.9 $\pm$ 14.6 \\
\midrule

\multicolumn{11}{l}{\emph{Operator splitting}} \\
\quad Reversed order (D $\rightarrow$ R)
& 10.4 $\pm$ 8.41
& 0.87 $\pm$ 0.16
& 3.30 $\pm$ 2.50
& 24.8 $\pm$ 11.0
& 9.45 $\pm$ 14.4
& 9.83 $\pm$ 9.67
& 5.91 $\pm$ 3.54
& \textbf{0.00 $\pm$ 0.00}
& 0.84 $\pm$ 2.81
& 0.19 $\pm$ 0.23 \\
\midrule

\multicolumn{11}{l}{\emph{Training}} \\
\quad w/o $\mathcal{L}_{\mathrm{phys}}$
& 7.49 $\pm$ 5.06
& 0.91 $\pm$ 0.09
& 2.20 $\pm$ 1.39
& 19.4 $\pm$ 9.1
& 5.23 $\pm$ 7.07
& 5.62 $\pm$ 4.04
& 4.55 $\pm$ 2.71
& \textbf{0.00 $\pm$ 0.00}
& 0.38 $\pm$ 0.60
& $<$ 0.01 \\
\quad w/o Stage C
& 10.0 $\pm$ 3.09
& 0.88 $\pm$ 0.06
& 5.02 $\pm$ 0.81
& --
& 3.93 $\pm$ 6.52
& 14.6 $\pm$ 1.31
& 16.4 $\pm$ 0.78
& --
& \textbf{0.00 $\pm$ 0.00}
& 100 $\pm$ 0.00 \\
\midrule

\textbf{PhysCARD (full)}
& \textbf{4.97 $\pm$ 2.41}
& \textbf{0.95 $\pm$ 0.04}
& \textbf{1.83 $\pm$ 1.09}
& \textbf{16.4 $\pm$ 8.9}
& \textbf{1.45 $\pm$ 0.85}
& \textbf{4.23 $\pm$ 2.76}
& 4.07 $\pm$ 3.08
& \textbf{0.00 $\pm$ 0.00}
& 0.02 $\pm$ 0.06
& $<$ 0.01 \\
\bottomrule
\end{tabular}
\end{adjustbox}
\end{table*}

We evaluated the contribution of each PhysCARD component by individually removing or modifying the reaction module, diffusion module, operator splitting, and training strategy (Table~\ref{tab:ablation}). The ablations revealed distinct effects on activation and repolarization dynamics. For repolarization, removing the rest anchor increased the repolarization-window $V_m$ RMSE from 16.4 to 39.8\,mV and the APD error from 4.07 to 15.90\,ms, indicating that small deviations from the resting equilibrium accumulate during autoregressive rollout. Similarly, omitting Stage~C increased the APD error to 16.37\,ms, with errors exceeding 10\,ms at all nodes, highlighting the importance of adapting pretrained cellular kinetics to coupled tissue dynamics. In contrast, activation accuracy was particularly sensitive to operator splitting and physics-guided learning. Reversing the reaction-then-diffusion order increased the LAT error from 1.45 to 9.45\,ms and the $V_m$ RMSE from 4.97 to 10.44\,mV, as diffusion errors near the excitation threshold can be amplified by nonlinear reaction kinetics. Removing the physics loss $\mathcal{L}_{\mathrm{phys}}$ likewise increased the LAT error to 5.23\,ms, demonstrating the importance of physically consistent predictions under perturbed states. Interestingly, replacing the conductance head with a per-edge MLP or removing its physical constraints had a comparatively modest effect on rollout accuracy, increasing the $V_m$ RMSE from 4.97 to 5.46 and 5.60\,mV, respectively. This suggests that accurate trajectories alone may not fully reflect the physical validity of the learned diffusion operator.

\begin{figure}[t]
    \centering
    \includegraphics[width=0.5\textwidth]{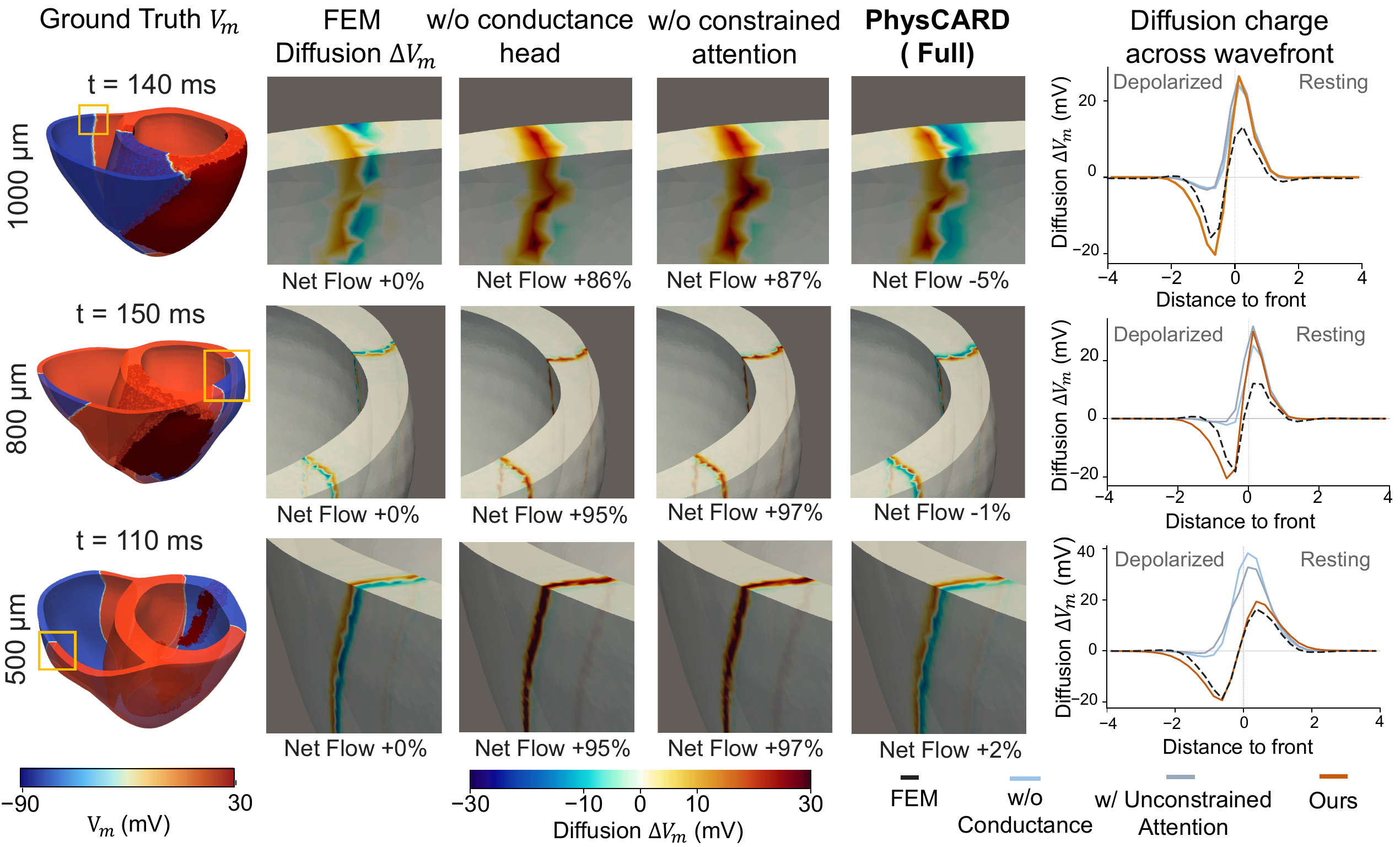}
        \vspace{-5mm}
    \caption{Comparison of FEM and learned diffusion-induced potential changes ($\Delta V_m$) at cardiac wavefronts across three mesh resolutions.}
    \label{fig:diffusion_flow}
\end{figure}

To further investigate this distinction, we isolated the diffusion module and compared its one-step potential updates against finite-element method (FEM) diffusion (Table~\ref{tab:diffusion_ablation}, Fig.~\ref{fig:diffusion_flow}). As shown in Fig.~\ref{fig:diffusion_flow}, FEM diffusion exhibits a bipolar pattern across the wavefront, redistributing potential from depolarized tissue behind the front to resting tissue ahead of it. PhysCARD closely reproduced this pattern, achieving a cosine similarity of 0.88, compared with 0.62 and 0.52 for the two ablated variants. Its symmetric, non-negative conductances ensured exact pairwise flux balance without violations of the discrete maximum principle. The volume-weighted net-charge metric showed a residual of $-7.5$\,\% across all evaluated states, decreasing from $-11.8$ to $-1.9$\,\% as the mesh resolution increased from 1000 to 500\,\textmu m. In contrast, both ablated variants exhibited predominantly monopolar patterns, adding potential ahead of the wavefront without a corresponding reduction behind it. Their net-charge imbalances reached 47.6\,\% and 42.5\,\%, respectively, with pairwise flux imbalances of 36.1\,\% and 70.4\,\%. These findings indicate that unconstrained diffusion modules may achieve competitive prediction accuracy by introducing artificial depolarization that compensates for errors elsewhere in the model, rather than learning physically valid diffusion. Notably, unconstrained attention showed lower agreement with FEM than the per-edge MLP, suggesting that the principal benefit arises from the imposed physical constraints rather than attention alone. 

\begin{table}[t]
\centering
\caption{One-step diffusion update compared with the FEM diffusion map.
Reaction disabled. Pattern agreement: cosine to the FEM increment; Net charge: $(P-N)/(P+N)$ of added and removed charge (0\,\% for pure diffusion). FEM: finite-element method.}
\label{tab:diffusion_ablation}

\footnotesize
\setlength{\tabcolsep}{2.5pt}
\renewcommand{\arraystretch}{1.12}

\begin{adjustbox}{max width=\columnwidth}
\begin{tabular}{@{}lccc@{}}
\toprule
\multirow{2}{*}{Metric}
& w/o conductance & w/o constrained & \multirow{2}{*}{\textbf{Ours}} \\
& head & attention & \\
% Metric
% & \shortstack{w/o conductance\\head}
% & \shortstack{Unconstrained\\attention}
% & \textbf{Ours} \\
\midrule

\multicolumn{4}{@{}l}{\emph{Agreement with FEM}} \\
\quad Pattern agreement (cos) $\uparrow$
& 0.62 $\pm$ 0.05 & 0.52 $\pm$ 0.06 & \textbf{0.88 $\pm$ 0.05} \\
\quad Front sign agreement (\%) $\uparrow$
& 88.4 $\pm$ 6.25 & 84.9 $\pm$ 5.23 & \textbf{96.6 $\pm$ 1.08} \\
\quad Net charge (\%) $\rightarrow 0$
& 47.6 $\pm$ 21.5 & 42.5 $\pm$ 10.2 & \textbf{$-$7.51 $\pm$ 4.00} \\

\midrule

\multicolumn{4}{@{}l}{\emph{Structural properties}} \\
\quad Flux imbalance (\%) $\downarrow$
& 36.1 $\pm$ 9.05 & 70.4 $\pm$ 5.00 & \textbf{0.00 $\pm$ 0.00} \\
\quad Max-principle viol.\ (\%) $\downarrow$
& 0.14 $\pm$ 0.08 & 0.66 $\pm$ 0.36 & \textbf{0.00 $\pm$ 0.00} \\
\quad Edges with $\alpha<0$ (\%) $\downarrow$
& 3.44 $\pm$ 2.35 & 0.57 $\pm$ 0.43 & \textbf{0.00 $\pm$ 0.00} \\
\quad Asymmetry $|\alpha_{ij}-\alpha_{ji}|$ (\%) $\downarrow$
& 64.3 $\pm$ 15.2 & 31.7 $\pm$ 4.92 & \textbf{0.00 $\pm$ 0.00} \\
\bottomrule
\end{tabular}
\end{adjustbox}
\end{table}

\subsection{Electrophysiology Fidelity}\label{sec:exp:ep_fidelity}
%---------------------------------------------------------
\begin{figure*}[!t]
\centering
\includegraphics[width=1\textwidth]{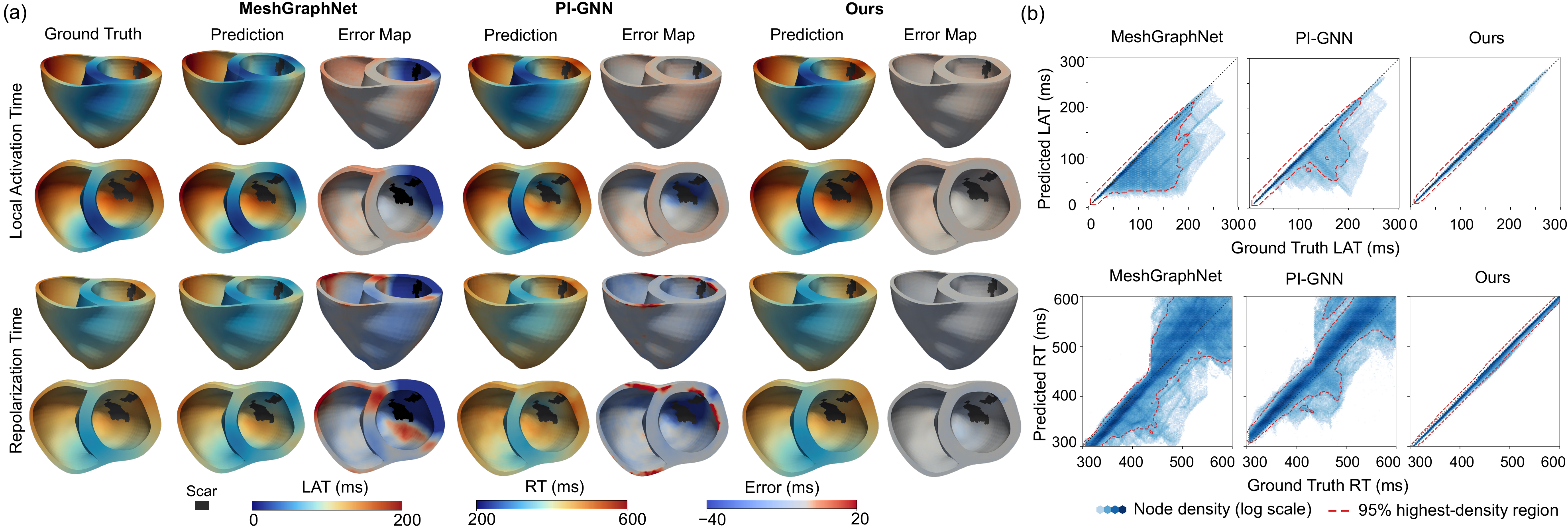}
\caption{
(a)~Local activation time and repolarization time on an unseen trajectory, with the corresponding error map.
(b)~Correlation between predicted versus ground-truth LAT and RT on the 20 test trajectories at 1000 and 800\,\textmu m; dotted line:
identity; red dashed line: 95\,\% highest-density region.}
\label{fig:lat_rt_map}
\end{figure*}

\begin{table}[!t]
\centering
\caption{Activation, repolarization and plausibility markers on the in-distribution test set.
$r$: Pearson correlation between predicted and ground-truth maps; AT90: time at which 90\,\% of the
myocardium is activated.} % PI-GNN w/o phys: PI-GNN without physics loss;
\label{tab:ep_markers}
\footnotesize
\setlength{\tabcolsep}{3pt}
\renewcommand{\arraystretch}{1.15}
\resizebox{\columnwidth}{!}{%
\begin{tabular}{l cc cc cc}
% \toprule
% & \multicolumn{2}{c}{Activation} & \multicolumn{2}{c}{Repolarization} & \multicolumn{2}{c}{Plausibility (\%) $\downarrow$} \\
% \cmidrule(lr){2-3}\cmidrule(lr){4-5}\cmidrule(lr){6-7}
% Method
% & \shortstack{LAT\\$r$ $\uparrow$} & \shortstack{AT90 err.\\(ms) $\downarrow$}
% & \shortstack{RT\\$r$ $\uparrow$} & \shortstack{APD bias\\(ms) $\rightarrow 0$}
% & \shortstack{Missed\\repol.} & \shortstack{Re-\\excitation} \\
% \midrule
\toprule
\multirow{3}{*}{Method}
& \multicolumn{2}{c}{Activation}
& \multicolumn{2}{c}{Repolarization}
& \multicolumn{2}{c}{Plausibility (\%) $\downarrow$} \\
\cmidrule(lr){2-3}\cmidrule(lr){4-5}\cmidrule(lr){6-7}
& \shortstack{LAT\\$r$ $\uparrow$}
& \shortstack{AT90 err.\\(ms) $\downarrow$}
& \shortstack{RT\\$r$ $\uparrow$}
& \shortstack{APD bias\\(ms) $\rightarrow 0$}
& \shortstack{Missed\\repol.}
& \shortstack{Re-\\excitation} \\
\midrule

MeshGraphNet
& 0.77 $\pm$ 0.16 & 31.4 $\pm$ 20.9
& 0.64 $\pm$ 0.30 & 14.9 $\pm$ 24.5
& 6.24 $\pm$ 6.82 & 35.7 $\pm$ 29.2 \\
% PI-GNN w/o phys
% & 0.85 $\pm$ 0.16 & 29.8 $\pm$ 24.7
% & 0.80 $\pm$ 0.20 & $-$14.9 $\pm$ 16.6
% & 0.03 $\pm$ 0.04 & 55.5 $\pm$ 40.7 \\
PI-GNN
& 0.91 $\pm$ 0.15 & 16.2 $\pm$ 20.3
& 0.82 $\pm$ 0.15 & 23.3 $\pm$ 8.64
& 11.4 $\pm$ 11.6 & 39.7 $\pm$ 30.7 \\
\textbf{Ours}
& \textbf{1.00 $\pm$ 0.00} & \textbf{1.56 $\pm$ 1.14}
& \textbf{1.00 $\pm$ 0.00} & \textbf{$-$3.21 $\pm$ 3.19}
& \textbf{0.02 $\pm$ 0.04} & \textbf{0.00 $\pm$ 0.00} \\
\bottomrule
\end{tabular}%
}
\end{table}

We further evaluated the fidelity of activation and repolarization patterns by comparing PhysCARD with MeshGraphNet~\citep{conf/pfaff2020} and PI-GNN~\citep{conf/MICCAI/tran2026}, which represent three levels of physical integration: data-driven message passing, physics-informed loss regularization, and physics-encoded reaction-diffusion operators. PhysCARD reproduced the activation sequence almost exactly, with a LAT correlation of $r=1.00$, an AT90 error of 1.56\,ms (Table~\ref{tab:ep_markers}), and LAT errors within about $\pm$4\,ms for 95\,\% of nodes, as presented in Fig.~\ref{fig:lat_rt_map} (b). Both other baselines, in contrast, activated the tissue too early, with mean LAT biases of $-19.1$ and $-7.73$\,ms, and the fan-shaped scatter shows that this bias grew with the distance travelled by the wavefront. This pattern indicates a conduction velocity that is slightly overestimated at each step and accumulates over the rollout. Physics-informed training reduced the drift, lowering the LAT MAE from 19.9 to 8.23\,ms (Table~\ref{tab:main}) and the AT90 error from 31.4 to 16.2\,ms, but did not remove it, whereas the constrained diffusion operator of PhysCARD lowered both errors by a further factor of 7 to 10. 

The contrast was larger in repolarization, where physics-informed training improved the spatial pattern but not the timing. PI-GNN raised the RT correlation of MeshGraphNet from 0.642 to 0.824, yet its RT MAE barely changed (27.9 vs.\
24.3\,ms), because it prolonged the action potential in every test trajectory, by 23.3\,ms on average. The smaller APD bias of MeshGraphNet (14.9\,ms) does not reflect better calibration: its errors had opposite signs at the two resolutions and partly cancelled, as its APD MAE of 25.5\,ms shows. Both baselines also left 6.24 and 11.4\,\% of nodes unrepolarized and re-excited 35.7 and 39.7\,\% of them, an artifact absent from the ground truth that downstream analyses could misread as arrhythmic activity. PhysCARD, in contrast, missed only 0.02\,\% of repolarizations and produced no re-excitation, and its action potentials followed the ground truth through the return to rest (Fig.~\ref{fig:qualitative}). This suggests that a physics loss constrains how the wavefront spreads but leaves recovery to be inferred from data, whereas the rest-anchored reaction module of PhysCARD ties recovery to the cellular kinetics throughout the rollout.

The remaining repolarization error of PhysCARD took its least harmful form, a small and uniform APD shortening. Its APD bias of $-3.21$\,ms (Table~\ref{tab:ep_markers}) averaged a near-zero error at 1000\,\textmu m with a shortening of 6.39\,ms at 800\,\textmu m, which reached 7.01\,ms at the unseen 500\,\textmu m resolution (Table~\ref{tab:runtime}). The shortening varied by only 0.04\,ms across trajectories, and the narrow RT limits of agreement ($\pm$6.50\,ms, Fig.~\ref{fig:lat_rt_map} b) show that it was also nearly constant across nodes, so it shifted repolarization earlier while the RT correlation remained 1.00. This resolution dependence may arise from interactions between the learned reaction dynamics and the mesh discretization, suggesting that resolution-aware calibration could further improve absolute repolarization timing.

\subsection{Physics Consistency}\label{sec:exp:phys_consistency}
%---------------------------------------------------------
\begin{figure}[t]
\centering
\includegraphics[width=\columnwidth]{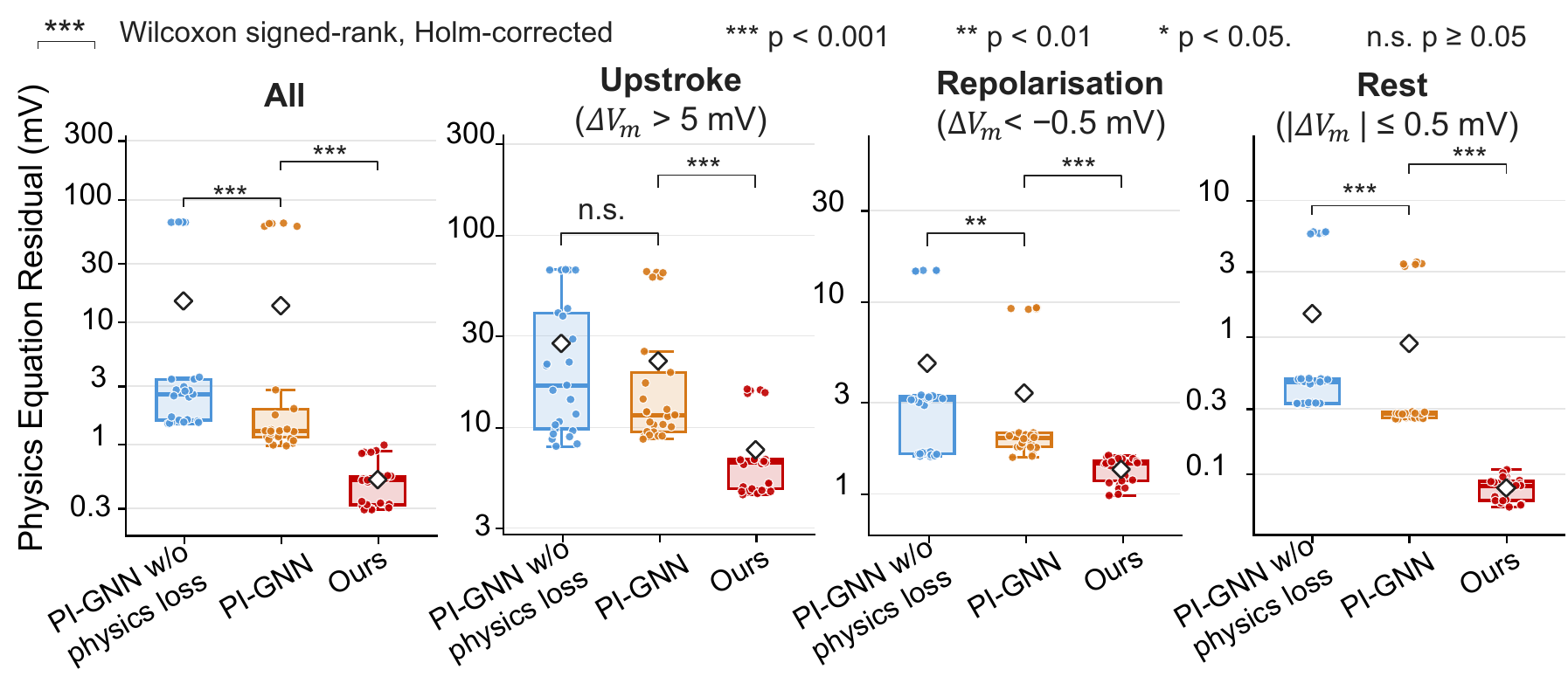}
\vspace{-9mm}
\caption{Physics equation residual per test anatomy over the whole heart and per action-potential region
($\diamond$: mean).}
\label{fig:physics_residual}
\end{figure}

To assess physical consistency beyond prediction accuracy, we evaluated the agreement between autoregressive predictions and the governing reaction-diffusion equations. At each rollout step, the solver-consistency residual was computed by comparing the model prediction with a reference numerical update from the same predicted state. We compared three levels of physical integration: data-driven PI-GNN without physics loss, PI-GNN with reaction-diffusion regularization~\citep{conf/MICCAI/tran2026}, and PhysCARD with physics-encoded operators. As shown in Fig.~\ref{fig:physics_residual}, the median residual decreased progressively from 2.6 to 1.3 and 0.52\,mV, with consistent improvements across all 25 test cases. The two baselines also exhibited substantially higher mean residuals (14.9 and 13.6\,mV), reflecting occasional large deviations from the governing dynamics during rollout.
The phase-specific analysis further revealed differences between loss-based and architecture-level physical constraints. Adding the physics loss reduced residuals at rest (0.47 to 0.28\,mV) and during repolarization (3.1 to 2.0\,mV), but its improvement during the rapid upstroke was not significant (16.6 to 11.6\,mV, $p=0.11$). In contrast, PhysCARD consistently reduced residuals across all phases, reaching 0.08\,mV at rest and 6.6\,mV during the upstroke. These improvements reflect the complementary roles of the rest-anchored reaction module and physics-constrained diffusion operator, which incorporate physical properties directly into the state update rather than relying solely on loss regularization.

\begin{figure}[t]
\centering
\includegraphics[width=\columnwidth]{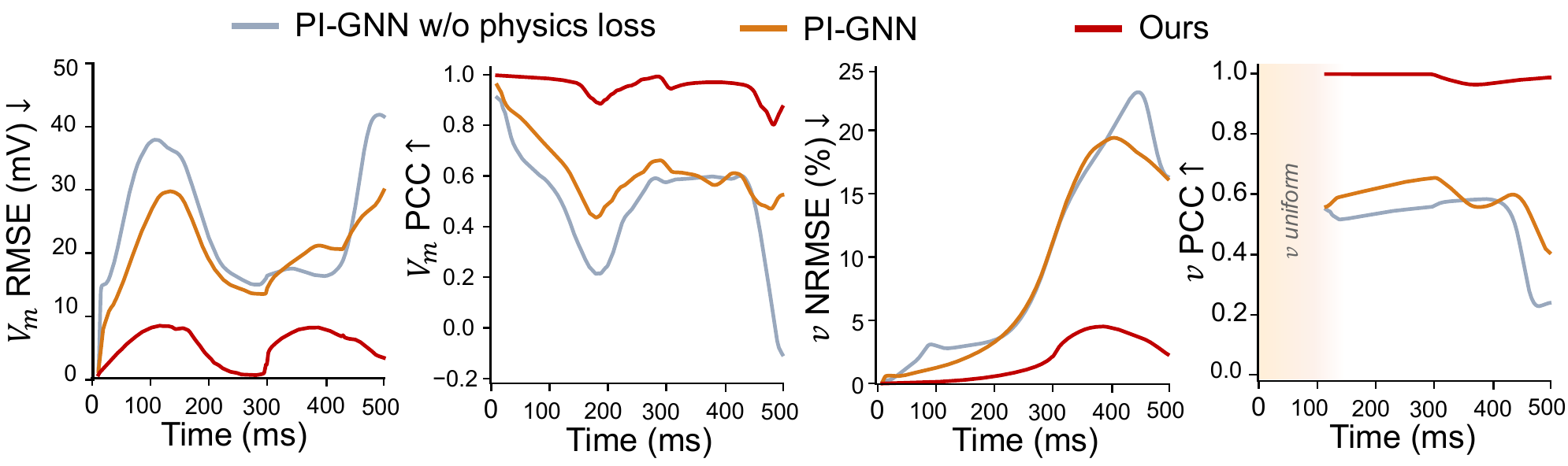}
\vspace{-8mm}
\caption{Rollout error over time, mean over the test set. Shaded: $v$ spatially uniform, PCC undefined.}
\label{fig:physics_contribution}
\end{figure}

The impact of these differences became more pronounced during long-horizon rollout (Fig.~\ref{fig:physics_contribution}). The peak $V_m$ RMSE during ventricular activation reached 37.7 and 29.6\,mV for the two baselines, compared with 8.3\,mV for PhysCARD. Baseline errors increased again during late repolarization, accompanied by accumulated errors in the recovery variable $v$, which reached 17-20\,\% by the end of the rollout, compared with 3.1\,\% for PhysCARD. Such drift in recovery dynamics contributes to the repolarization inaccuracies observed in Sec.~\ref{sec:exp:ep_fidelity}. Together, these results demonstrate that embedding reaction-diffusion structure into the model provides more consistent EP dynamics across action-potential phases and substantially reduces error accumulation during autoregressive simulation.

\subsection{Cross-Resolution Scalability and Computational Efficiency}\label{sec:exp:scalability}
%---------------------------------------------------------
\begin{table*}[t]
\centering
\caption{Computational cost and accuracy of PhysCARD at three mesh resolutions (500\,\textmu m is unseen in training). Runtime: wall-clock time to advance one frame (1\,ms of simulated
time), with openCARP using its GPU linear solver and PhysCARD on the same GPU; speed-up: ratio of the
two. }
\label{tab:runtime}
\footnotesize
\setlength{\tabcolsep}{4pt}
\renewcommand{\arraystretch}{1.15}
\begin{adjustbox}{max width=\textwidth}
\begin{tabular}{l cc ccc cccc}
\toprule
& \multicolumn{2}{c}{Mesh}
& \multicolumn{3}{c}{Runtime per frame}
& \multicolumn{4}{c}{Accuracy (full free rollout)} \\
\cmidrule(lr){2-3}\cmidrule(lr){4-6}\cmidrule(lr){7-10}
Resolution & N. of Nodes & N. of Edges
& openCARP (s) & PhysCARD (ms) & Speed-up
& $V_m$ RMSE (mV) & $v$ NRMSE (\%) & LAT (ms) & APD (ms) \\
\midrule
1000\,\textmu m & 148 $\pm$ 26.1K & 1.82 $\pm$ 0.329M
& 1.21 $\pm$ 0.103 & 7.20 $\pm$ 1.46 & 168$\times$
& 2.57 $\pm$ 0.953 & 0.619 $\pm$ 0.200 & 1.12 $\pm$ 0.652 & 0.324 $\pm$ 0.006 \\
800\,\textmu m & 312 $\pm$ 51.6K & 4.04 $\pm$ 0.670M
& 1.71 $\pm$ 0.160 & 14.7 $\pm$ 2.61 & 116$\times$
& 6.47 $\pm$ 1.11 & 2.83 $\pm$ 0.308 & 1.35 $\pm$ 0.737 & 6.40 $\pm$ 0.043 \\
500\,\textmu m & 1.40 $\pm$ 0.232M & 17.3 $\pm$ 2.89M
& 4.96 $\pm$ 1.27 & 46.7 $\pm$ 9.47 & 106$\times$
& 7.20 $\pm$ 1.27 & 2.60 $\pm$ 0.450 & 2.36 $\pm$ 0.735 & 7.02 $\pm$ 0.015 \\
\bottomrule
\end{tabular}
\end{adjustbox}
\end{table*}

We evaluated the scalability and computational efficiency of PhysCARD across three mesh resolutions (Table~\ref{tab:runtime}). The 500\,\textmu m meshes provide a particularly challenging setting, with an unseen spatial resolution and approximately 4.5 times more nodes than the 800\,\textmu m meshes. Despite this increase in complexity, PhysCARD maintained comparable accuracy, with the rollout $V_m$ RMSE increasing only from 6.47 to 7.20\,mV, the $v$ NRMSE remaining similar (2.83\,\% vs.\ 2.60\,\%), and the LAT error remaining below 2.4\,ms. This scalability is supported by the mesh-independent reaction module and the locally defined, physics-constrained diffusion operator, which operate on variable mesh geometries without retraining. The remaining differences were mainly observed in repolarization, where APD errors increased from 0.32\,ms at 1000\,\textmu m to 6.40 and 7.02\,ms at 800 and 500\,\textmu m, respectively. As discussed in Sec.~\ref{sec:exp:ep_fidelity}, these errors primarily reflect a systematic shift in repolarization timing rather than a loss of spatial propagation accuracy.

PhysCARD also achieved substantial computational acceleration across all resolutions. The inference time per 1\,ms simulation frame was 7.20, 14.7, and 46.7\,ms at 1000, 800, and 500\,\textmu m, respectively, compared with 1.21, 1.71, and 4.96\,s for OpenCARP. This corresponds to speed-ups of 168$\times$, 116$\times$, and 106$\times$, respectively, averaging 130$\times$ across resolutions. For a 600\,ms cardiac simulation, PhysCARD reduced the runtime from approximately 12\,min to 4\,s at 1000\,\textmu m, and from 50\,min to 28\,s at 500\,\textmu m. 
Notably, the cumulative savings are particularly relevant to CDTs, where parameter calibration may require tens of thousands of simulations~\citep{journal/MedIA/camps2025}. 
\section{Discussion and Conclusion}
\label{sec:discussion}
% =========================================================

In this study, we developed PhysCARD to address the challenges of accurate and efficient long-horizon EP simulation for CDTs. The key methodological contribution is a physics-grounded reaction-diffusion framework that explicitly separates cellular kinetics from spatial electrical propagation. By combining a rest-anchored reaction flow, physics-encoded conductance diffusion, and a staged learning strategy, PhysCARD incorporates essential physical properties directly into the neural simulation process. We conducted comprehensive experiments on patient-specific biventricular geometries with myocardial scar to evaluate different aspects of the framework. Comparisons with existing neural surrogates demonstrated improvements in prediction accuracy and long-horizon stability (Sec.~\ref{sec:exp:comparison}), while ablation studies examined the contributions of individual components and the physical validity of learned diffusion (Sec.~\ref{sec:exp:ablation}). Further analyses investigated EP fidelity (Sec.~\ref{sec:exp:ep_fidelity}), consistency with the governing equations (Sec.~\ref{sec:exp:phys_consistency}), and scalability and computational efficiency across mesh resolutions (Sec.~\ref{sec:exp:scalability}). Collectively, these findings highlight the importance of embedding physical structure into neural simulators, rather than relying solely on physics-informed loss regularization. In particular, the diffusion ablation revealed that comparable prediction accuracy can arise from physically inconsistent operators, underscoring the need to evaluate both predictive performance and the underlying physical mechanisms.

Despite these improvements, several limitations remain. First, PhysCARD currently relies on the simplified Aliev-Panfilov model, which does not explicitly represent detailed ionic currents or complex arrhythmogenic mechanisms. Its applicability to more sophisticated cellular models and reentrant arrhythmias therefore remains to be established. Second, although PhysCARD demonstrated cross-resolution scalability, the systematic shortening of action potential duration at finer resolutions suggests that the learned reaction dynamics may remain sensitive to spatial discretization. Third, the autoregressive formulation requires sequential reaction-diffusion updates, introducing a computational cost that increases with simulation duration. Finally, the framework was evaluated against numerical simulations derived from a relatively small, single-centre clinical cohort. Its performance under broader physiological conditions and agreement with measured intracardiac electrical activity require further validation.

Future work will extend the framework to more detailed ionic models, such as the ten~Tusscher and O'Hara-Rudy formulations, and investigate complex excitation patterns, including reentrant arrhythmias. We will also explore resolution-aware reaction modelling and larger integration steps to improve scalability and computational efficiency, alongside validation using clinical electroanatomical mapping data. Ultimately, PhysCARD provides a foundation for developing cardiac EP surrogates that combine computational efficiency with physically interpretable dynamics, supporting the advancement of real-time CDTs for personalized healthcare.

\bibliographystyle{model2-names}
\biboptions{authoryear}
\bibliography{A_refs}

@article{fresca2020deep,
  title={Deep learning-based reduced order models in cardiac electrophysiology},
  author={Fresca, Stefania and Manzoni, Andrea and Ded{\`e}, Luca and Quarteroni, Alfio},
  journal={PloS one},
  volume={15},
  number={10},
  pages={e0239416},
  year={2020},
  publisher={Public Library of Science San Francisco, CA USA}
}

@article{journal/arxiv/lyu2026,
  title={Personalized 3D Myocardial Infarct Geometry Reconstruction from Cine MRI for Cardiac Digital Twins},
  author={Lyu, Yilin and Chan, Mark YY and Sia, Ching-Hui and Li, Lei},
  journal={arXiv preprint arXiv:2606.01808},
  year={2026}
}

@article{journal/NEJM/chrispin2026,
  title={Digital twin-guided ablation for ventricular tachycardia},
  author={Chrispin, Jonathan and Prakosa, Adityo and Kholmovski, Eugene and Kolandaivelu, Aravindan and Barcelon, Amanda and Aronis, Konstantinos N and Berger, Ronald D and Calkins, Hugh and Trayanova, Natalia},
  journal={The New England journal of medicine},
  volume={394},
  number={13},
  pages={1345},
  year={2026}
}

@article{journal/NC/arevalo2016,
  title={Arrhythmia risk stratification of patients after myocardial infarction using personalized heart models},
  author={Arevalo, Hermenegild J and Vadakkumpadan, Fijoy and Guallar, Eliseo and Jebb, Alexander and Malamas, Peter and Wu, Katherine C and Trayanova, Natalia A},
  journal={Nature communications},
  volume={7},
  number={1},
  pages={11437},
  year={2016},
  publisher={Nature Publishing Group UK, London}
}

@article{journal/EHJ/corral2020,
  title={The ‘Digital Twin’ to enable the vision of precision cardiology},
  author={Corral-Acero, Jorge and Margara, Francesca and Marciniak, Maciej and Rodero, Cristobal and Loncaric, Filip and Feng, Yingjing and Gilbert, Andrew and Fernandes, Joao F and Bukhari, Hassaan A and Wajdan, Ali and Martinez, Manuel Villegas and Santos, Mariana Sousa and Shamohammdi, Mehrdad and Luo, Hongxing and Westphal, Philip and Leeson, Paul and DiAchille, Paolo and Gurev, Viatcheslav and Mayr, Manuel and Geris, Liesbet and Pathmanathan, Pras and Morrison, Tina and Cornelussen, Richard and Prinzen, Frits and Delhaas, Tammo and Doltra, Ada and Sitges, Marta and Vigmond, Edward J and Zacur, Ernesto and Grau, Vicente and Rodriguez, Blanca and Remme, Espen W and Niederer, Steven and Mortier, Peter and McLeod, Kristin and Potse, Mark and Pueyo, Esther and Bueno-Orovio, Alfonso and Lamata, Pablo},
  journal={European Heart Journal},
  volume={41},
  number={48},
  pages={4556--4564},
  year={2020},
  publisher={Oxford University Press}
}

@article{fresca2021pod,
  title={POD-enhanced deep learning-based reduced order models for the real-time simulation of cardiac electrophysiology in the left atrium},
  author={Fresca, Stefania and Manzoni, Andrea and Ded{\`e}, Luca and Quarteroni, Alfio},
  journal={Frontiers in physiology},
  volume={12},
  pages={679076},
  year={2021},
  publisher={Frontiers Media SA}
}

@article{herrero2022eppinn,
  title={EP-PINNs: Cardiac electrophysiology characterisation using physics-informed neural networks},
  author={Herrero Martin, Clara and Oved, Alon and Chowdhury, Rasheda A and Ullmann, Elisabeth and Peters, Nicholas S and Bharath, Anil A and Varela, Marta},
  journal={Frontiers in Cardiovascular Medicine},
  volume={8},
  pages={768419},
  year={2022},
  publisher={Frontiers Media SA}
}

@inproceedings{chiu2024physics,
  title={Physics-Informed Neural Networks can accurately model cardiac electrophysiology in 3D geometries and fibrillatory conditions},
  author={Chiu, Ching-En and Roy, Aditi and Cechnicka, Sarah and Gupta, Ashvin and Pinto, Arieh Levy and Galazis, Christoforos and Christensen, Kim and Mandic, Danilo and Varela, Marta},
  booktitle={International Workshop on Statistical Atlases and Computational Models of the Heart},
  pages={98--109},
  year={2024},
  organization={Springer}
}

@article{naghavi2026rapid,
  title={Rapid prediction of cardiac activation in the left ventricle with geometric deep learning: a step towards cardiac resynchronization therapy planning},
  author={Naghavi, Ehsan and Wang, Haifeng and Ziaei-Rad, Vahid and Guccione, Julius and Kassab, Ghassan and Boddeti, Vishnu and Baek, Seungik and Lee, Lik-Chuan},
  journal={npj Digital Medicine},
  year={2026},
  publisher={Nature Publishing Group UK London}
}

@inproceedings{conf/ICFIM/morier2025lagata,
  title={Learning Cardiac Electrophysiology with Graph Neural Networks for Fast Data-driven Personalised Predictions},
  author={Morier, Ma{\"e}lis and Rodr{\'\i}guez Padilla, Jairo and Gallinari, Patrick and Sermesant, Maxime},
  booktitle={International Conference on Functional Imaging and Modeling of the Heart},
  pages={138--150},
  year={2025},
  organization={Springer}
}

@inproceedings{dong2023deep,
  title={Deep-Learning-Assisted Cardiac Electrophysiology Simulation},
  author={Dong, Weixuan and Li, Yifu and Zhu, Rui},
  booktitle={2023 Winter Simulation Conference (WSC)},
  pages={758--769},
  year={2023},
  organization={IEEE}
}

@inproceedings{ayed2019ep,
  title={Ep-net: Learning cardiac electrophysiology models for physiology-based constraints in data-driven predictions},
  author={Ayed, Ibrahim and Cedilnik, Nicolas and Gallinari, Patrick and Sermesant, Maxime},
  booktitle={International Conference on Functional Imaging and Modeling of the Heart},
  pages={55--63},
  year={2019},
  organization={Springer}
}

@inproceedings{kashtanova2021ep,
  title={EP-Net 2.0: Out-of-domain generalisation for deep learning models of cardiac electrophysiology},
  author={Kashtanova, Victoriya and Ayed, Ibrahim and Cedilnik, Nicolas and Gallinari, Patrick and Sermesant, Maxime},
  booktitle={International Conference on Functional Imaging and Modeling of the Heart},
  pages={482--492},
  year={2021},
  organization={Springer}
}

@article{martinez2026full,
  title={Full-field surrogate modeling of cardiac electrophysiology encoding geometric variability},
  author={Martinez, Elena Sabdy and Moscoloni, Beatrice and Salvador, Matteo and Kong, Fanwei and Peirlinck, Mathias and Marsden, Alison Lesley},
  journal={Computer Methods in Applied Mechanics and Engineering},
  volume={448},
  pages={118444},
  year={2026},
  publisher={Elsevier}
}

@article{zhou2025unified,
  title={A unified framework for geometry-independent operator learning in cardiac electrophysiology simulations},
  author={Zhou, Bei and Corrado, Cesare and Qian, Shuang and Balmus, Maximilian and Lee, Angela WC and Rodero, Cristobal and Roney, Caroline and Gotte, Marco JW and Hopman, Luuk HGA and Plank, Gernot and others},
  journal={arXiv preprint arXiv:2512.01702},
  year={2025}
}

@article{ziarelli2026learning,
  title={Learning cardiac activation and repolarization times with operator learning},
  author={Ziarelli, Giovanni and Centofanti, Edoardo and Parolini, Nicola and Scacchi, Simone and Verani, Marco and Pavarino, Luca F},
  journal={PLOS Computational Biology},
  volume={22},
  number={1},
  pages={e1013920},
  year={2026},
  publisher={Public Library of Science San Francisco, CA USA}
}

@article{journal/CM/vicente2026,
  title={Volumetric non-invasive cardiac mapping for accessible global arrhythmia characterization},
  author={Vicente-Puig, Jorge and Chamorro-Servent, Judit and Zacur, Ernesto and Llorente-Lipe, In{\'e}s and Mart{\'\i}nez-P{\'e}rez, Marta and S{\'a}nchez, Jorge and Reventos-Presmanes, Jana and Roca-Luque, Ivo and Mont, Llu{\'\i}s and Atienza, Felipe and others},
  journal={Communications Medicine},
  year={2026},
  publisher={Nature Publishing Group UK London}
}

@article{journal/RiBME/li2024,
  title={Solving the inverse problem of electrocardiography for cardiac digital twins: A survey},
  author={Li, Lei and Camps, Julia and Rodriguez, Blanca and Grau, Vicente},
  journal={IEEE Reviews in Biomedical Engineering},
  volume={18},
  pages={316--336},
  year={2024},
  publisher={IEEE}
}

@article{journal/FiP/kalinin2019,
  title={Solving the inverse problem of electrocardiography on the endocardium using a single layer source},
  author={Kalinin, Alexander and Potyagaylo, Danila and Kalinin, Vitaly},
  journal={Frontiers in physiology},
  volume={10},
  pages={58},
  year={2019},
  publisher={Frontiers Media SA}
}

@article{journal/NHJ/cluitmans2015,
  title={Noninvasive reconstruction of cardiac electrical activity: update on current methods, applications and challenges},
  author={Cluitmans, Matthijs JM and Peeters, RLM and Westra, RL and Volders, PGA},
  journal={Netherlands Heart Journal},
  volume={23},
  number={6},
  pages={301--311},
  year={2015},
  publisher={Springer}
}

@article{journal/ABME/crozier2016,
  title={Image-based personalization of cardiac anatomy for coupled electromechanical modeling},
  author={Crozier, A and Augustin, CM and Neic, A and Prassl, AJ and Holler, M and Fastl, TE and Hennemuth, A and Bredies, K and Kuehne, T and Bishop, MJ and others},
  journal={Annals of biomedical engineering},
  volume={44},
  number={1},
  pages={58--70},
  year={2016},
  publisher={Springer}
}

@article{journal/nature/arevalo2016,
  title={Arrhythmia risk stratification of patients after myocardial infarction using personalized heart models},
  author={Arevalo, Hermenegild J and Vadakkumpadan, Fijoy and Guallar, Eliseo and Jebb, Alexander and Malamas, Peter and Wu, Katherine C and Trayanova, Natalia A},
  journal={Nature communications},
  volume={7},
  number={1},
  pages={11437},
  year={2016},
  publisher={Nature Publishing Group UK London}
}

@article{journal/SR/wang2019,
  title={Simulation of action potential propagation based on the ghost structure method},
  author={Wang, Yongheng and Cai, Li and Luo, Xiaoyu and Ying, Wenjun and Gao, Hao},
  journal={Scientific reports},
  volume={9},
  number={1},
  pages={10927},
  year={2019},
  publisher={Nature Publishing Group UK London}
}

@article{journal/physiol/vincent2015,
  title={High-order finite element methods for cardiac monodomain simulations Front},
  author={Vincent, KP and Gonzales, MJ and Gillette, AK and Villongco, CT and Pezzuto, S and Omens, JH and Holst, MJ and McCulloch, AD},
  journal={Physiol},
  volume={6},
  number={217},
  pages={9},
  year={2015}
}

@article{journal/CAM/roy2020,
  title={Analysis of time-stepping methods for the monodomain model: T. Roy et al.},
  author={Roy, Thomas and Bourgault, Yves and Pierre, Charles},
  journal={Computational and Applied Mathematics},
  volume={39},
  number={3},
  pages={230},
  year={2020},
  publisher={Springer}
}

@article{journal/niederer2020,
  title={Creation and application of virtual patient cohorts of heart models},
  author={Niederer, Steven A and Aboelkassem, Yasser and Cantwell, Chris D and Corrado, Cesare and Coveney, Sam and Cherry, Elizabeth M and Delhaas, Tammo and Fenton, Flavio H and Panfilov, Alexander V and Pathmanathan, Pras and others},
  journal={Philosophical transactions. Series A, Mathematical, physical, and engineering sciences},
  volume={378},
  number={2173},
  pages={20190558},
  year={2020}
}

@article{journal/raissi2019,
  title={Physics-informed neural networks: A deep learning framework for solving forward and inverse problems involving nonlinear partial differential equations},
  author={Raissi, Maziar and Perdikaris, Paris and Karniadakis, George E},
  journal={Journal of Computational physics},
  volume={378},
  pages={686--707},
  year={2019},
  publisher={Elsevier}
}

@article{journal/FiP/sahli2020,
  title={Physics-informed neural networks for cardiac activation mapping},
  author={Sahli Costabal, Francisco and Yang, Yibo and Perdikaris, Paris and Hurtado, Daniel E and Kuhl, Ellen},
  journal={Frontiers in Physics},
  volume={8},
  pages={42},
  year={2020},
  publisher={Frontiers Media SA}
}

@article{journal/IEEE/xie2022,
  title={Physics-constrained deep learning for robust inverse ecg modeling},
  author={Xie, Jianxin and Yao, Bing},
  journal={IEEE Transactions on Automation Science and Engineering},
  volume={20},
  number={1},
  pages={151--166},
  year={2022},
  publisher={IEEE}
}

@article{lydon2025,
  title={Physics-Informed Neural Operators for Cardiac Electrophysiology},
  author={Lydon, Hannah and Kazemi, Milad and Bishop, Martin and Paoletti, Nicola},
  journal={arXiv preprint arXiv:2511.08418},
  year={2025}
}

@inproceedings{conf/MICCAI/toloubidokhti2025,
  title={Meta-learning Physics-Informed Neural Networks for Personalized Cardiac Modeling},
  author={Toloubidokhti, Maryam and Missel, Ryan and Lian, Shichang and Wang, Linwei},
  booktitle={International Conference on Medical Image Computing and Computer-Assisted Intervention},
  pages={344--354},
  year={2025},
  organization={Springer}
}

@inproceedings{book/kashtanova2022,
  title={APHYN-EP: Physics-based deep learning framework to learn and forecast cardiac electrophysiology dynamics},
  author={Kashtanova, Victoriya and Pop, Mihaela and Ayed, Ibrahim and Gallinari, Patrick and Sermesant, Maxime},
  booktitle={International Workshop on Statistical Atlases and Computational Models of the Heart},
  pages={190--199},
  year={2022},
  organization={Springer}
}

@article{arxiv/jiang2024,
  title={Hyper-EP: Meta-learning hybrid personalized models for cardiac electrophysiology},
  author={Jiang, Xiajun and Vadhavkar, Sumeet and Ye, Yubo and Toloubidokhti, Maryam and Missel, Ryan and Wang, Linwei},
  journal={arXiv preprint arXiv:2403.15433},
  year={2024}
}

@inproceedings{conf/ICML/cao2023,
  title={Efficient learning of mesh-based physical simulation with bi-stride multi-scale graph neural network},
  author={Cao, Yadi and Chai, Menglei and Li, Minchen and Jiang, Chenfanfu},
  booktitle={International conference on machine learning},
  pages={3541--3558},
  year={2023},
  organization={PMLR}
}

@article{conf/pfaff2020,
  title={Learning mesh-based simulation with graph networks},
  author={Pfaff, Tobias and Fortunato, Meire and Sanchez-Gonzalez, Alvaro and Battaglia, Peter W},
  journal={arXiv preprint arXiv:2010.03409},
  year={2020}
}

@article{conf/fortunato2022,
  title={Multiscale meshgraphnets},
  author={Fortunato, Meire and Pfaff, Tobias and Wirnsberger, Peter and Pritzel, Alexander and Battaglia, Peter},
  journal={arXiv preprint arXiv:2210.00612},
  year={2022}
}

@article{conf/nabian2024,
  title={X-meshgraphnet: Scalable multi-scale graph neural networks for physics simulation},
  author={Nabian, Mohammad Amin and Liu, Chang and Ranade, Rishikesh and Choudhry, Sanjay},
  journal={arXiv preprint arXiv:2411.17164},
  year={2024}
}

@article{conf/iparraguirre2026,
  title={MeshGraphNet-Transformer: Scalable Mesh-based Learned Simulation for Solid Mechanics},
  author={Iparraguirre, Mikel M and Alfaro, Iciar and Gonzalez, David and Cueto, Elias},
  journal={arXiv preprint arXiv:2601.23177},
  year={2026}
}

@article{journal/CMAM/wurth2024,
  title={Physics-informed MeshGraphNets (PI-MGNs): Neural finite element solvers for non-stationary and nonlinear simulations on arbitrary meshes},
  author={W{\"u}rth, Tobias and Freymuth, Niklas and Zimmerling, Clemens and Neumann, Gerhard and K{\"a}rger, Luise},
  journal={Computer Methods in Applied Mechanics and Engineering},
  volume={429},
  pages={117102},
  year={2024},
  publisher={Elsevier}
}

@inproceedings{book/de2024,
  title={Finite volume informed graph neural network for myocardial perfusion simulation},
  author={de Chou, Raoul Sall{\'e} and Sinclair, Matthew and Lynch, Sabrina and Xiao, Nan and Najman, Laurent and Vignon-Clementel, Irene and Talbot, Hugues},
  booktitle={Medical Imaging with Deep Learning},
  year={2024}
}

@article{journal/ASC/xiang2024,
  title={Solving spatiotemporal partial differential equations with physics-informed graph neural network},
  author={Xiang, Zixue and Peng, Wei and Yao, Wen and Liu, Xu and Zhang, Xiaoya},
  journal={Applied Soft Computing},
  volume={155},
  pages={111437},
  year={2024},
  publisher={Elsevier}
}

@inproceedings{conf/AAAI/yang2026,
  title={Pegnet: A physics-embedded graph network for long-term stable multiphysics simulation},
  author={Yang, Can and Wang, Zhenzhong and Liu, Junyuan and Gong, Yunpeng and Jiang, Min},
  booktitle={Proceedings of the AAAI Conference on Artificial Intelligence},
  volume={40},
  number={32},
  pages={27494--27502},
  year={2026}
}

@inproceedings{conf/ICLR/zeng2025,
  title={PhyMPGN: Physics-encoded message passing graph network for spatiotemporal PDE systems},
  author={Zeng, Bocheng and Wang, Qi and Yan, Mengtao and Liu, Yang and Chengze, Ruizhi and Zhang, Yi and Liu, Hongsheng and Wang, Zidong and Sun, Hao},
  booktitle={International conference on learning representations},
  volume={2025},
  pages={74179--74206},
  year={2025}
}

@inproceedings{workshop/li2022,
  title={Deep computational model for the inference of ventricular activation properties},
  author={Li, Lei and Camps, Julia and Banerjee, Abhirup and Beetz, Marcel and Rodriguez, Blanca and Grau, Vicente},
  booktitle={International Workshop on Statistical Atlases and Computational Models of the Heart},
  pages={369--380},
  year={2022},
  organization={Springer}
}

@article{arxiv/yin2024,
  title={Dimon: Learning solution operators of partial differential equations on a diffeomorphic family of domains},
  author={Yin, Minglang and Charon, Nicolas and Brody, Ryan and Lu, Lu and Trayanova, Natalia and Maggioni, Mauro},
  journal={arXiv preprint arXiv:2402.07250},
  year={2024}
}

@inproceedings{conf/MICCAI/ye2023,
  title={A spatial-temporally adaptive PINN framework for 3D bi-ventricular electrophysiological simulations and parameter inference},
  author={Ye, Yubo and Liu, Huafeng and Jiang, Xiajun and Toloubidokhti, Maryam and Wang, Linwei},
  booktitle={International Conference on Medical Image Computing and Computer-Assisted Intervention},
  pages={163--172},
  year={2023},
  organization={Springer}
}

@article{journal/aliev1996,
  title={A simple two-variable model of cardiac excitation},
  author={Aliev, Rubin R and Panfilov, Alexander V},
  journal={Chaos, Solitons \& Fractals},
  volume={7},
  number={3},
  pages={293--301},
  year={1996},
  publisher={Elsevier}
}

@article{journal/JCP/lan2023dosnet,
  title={DOSnet as a non-black-box PDE solver: When deep learning meets operator splitting},
  author={Lan, Yuan and Li, Zhen and Sun, Jie and Xiang, Yang},
  journal={Journal of Computational Physics},
  volume={491},
  pages={112343},
  year={2023},
  publisher={Elsevier}
}

@inproceedings{conf/ICML/karlbauer2022,
  title={Composing partial differential equations with physics-aware neural networks},
  author={Karlbauer, Matthias and Praditia, Timothy and Otte, Sebastian and Oladyshkin, Sergey and Nowak, Wolfgang and Butz, Martin V},
  booktitle={International Conference on Machine Learning},
  pages={10773--10801},
  year={2022},
  organization={PMLR}
}

@article{conf/vadhavkar2026hapi,
  title={HAPI-EP: Towards Hybrid, Adaptive, and Predictive Digital Twins of Cardiac Electrophysiology},
  author={Vadhavkar, Sumeet and Jiang, Xiajun and Ye, Yubo and Toloubidokhti, Maryam and Wang, Linwei},
  journal={arXiv preprint arXiv:2606.15637},
  year={2026}
}

@article{journal/JCP/qin2019,
  title={Data driven governing equations approximation using deep neural networks},
  author={Qin, Tong and Wu, Kailiang and Xiu, Dongbin},
  journal={Journal of Computational Physics},
  volume={395},
  pages={620--635},
  year={2019},
  publisher={Elsevier}
}

@inproceedings{conf/MICCAI/tran2026,
  title={Physics-Informed Surrogate Model Using Graph Neural Network for Cardiac Electrophysiology},
  author={Tran, Vu Anh and Wang, Mengxiao and Lyu, Yilin and Chauhan, Jagmohan and Muthurangu, Vivek and Sia, Ching-Hui and Li, Lei},
  booktitle={International Conference on Medical Image Computing and Computer-Assisted Intervention},
  pages={407--417},
  year={2026},
  organization={Springer}
}

@article{journal/ANIPS/li2023gino,
  title={Geometry-informed neural operator for large-scale 3d pdes},
  author={Li, Zongyi and Kovachki, Nikola and Choy, Chris and Li, Boyi and Kossaifi, Jean and Otta, Shourya and Nabian, Mohammad Amin and Stadler, Maximilian and Hundt, Christian and Azizzadenesheli, Kamyar and others},
  journal={Advances in Neural Information Processing Systems},
  volume={36},
  pages={35836--35854},
  year={2023}
}

@inproceedings{conf/iclm/chamberlain2021,
  title={Grand: Graph neural diffusion},
  author={Chamberlain, Ben and Rowbottom, James and Gorinova, Maria I and Bronstein, Michael and Webb, Stefan and Rossi, Emanuele},
  booktitle={International conference on machine learning},
  pages={1407--1418},
  year={2021},
  organization={PMLR}
}

@article{journal/vaswani2017,
  title={Attention is all you need},
  author={Vaswani, Ashish and Shazeer, Noam and Parmar, Niki and Uszkoreit, Jakob and Jones, Llion and Gomez, Aidan N and Kaiser, {\L}ukasz and Polosukhin, Illia},
  journal={Advances in neural information processing systems},
  volume={30},
  year={2017}
}

@article{journal/MedIA/dillon2026,
  title={biv-me: Open-source software for generating time-varying biventricular meshes from cine cardiovascular magnetic resonance imaging with multi-cohort validation},
  author={Dillon, Joshua and Mauger, Charl{\`e}ne and Zhao, Debbie and Petersen, Steffen E and McCulloch, Andrew D and Young, Alistair A and Nash, Martyn P},
  journal={Medical Image Analysis},
  pages={104252},
  year={2026},
  publisher={Elsevier}
}

@inproceedings{conf/mauger2018,
  title={An iterative diffeomorphic algorithm for registration of subdivision surfaces: application to congenital heart disease},
  author={Mauger, Charlene and Gilbert, Kathleen and Suinesiaputra, Avan and Pontre, Beau and Omens, J and McCulloch, A and Young, A},
  booktitle={2018 40th annual international conference of the IEEE engineering in medicine and biology society (EMBC)},
  pages={596--599},
  year={2018},
  organization={IEEE}
}

@article{journal/MedIA/schuler2021,
  title={Cobiveco: Consistent biventricular coordinates for precise and intuitive description of position in the heart--with MATLAB implementation},
  author={Schuler, Steffen and Pilia, Nicolas and Potyagaylo, Danila and Loewe, Axel},
  journal={Medical Image Analysis},
  volume={74},
  pages={102247},
  year={2021},
  publisher={Elsevier}
}

@article{journal/doste2026,
  title={An automated computational pipeline for generating large-scale cohorts of patient-specific ventricular models in electromechanical in silico trials},
  author={Doste, Ruben and Camps, Julia and Wang, Zhinuo Jenny and Berg, Lucas Arantes and Holmes, Maxx and Smith, Hannah and Beetz, Marcel and Li, Lei and Banerjee, Abhirup and Grau, Vicente and others},
  journal={Computer Methods and Programs in Biomedicine},
  pages={109290},
  year={2026},
  publisher={Elsevier}
}

@article{journal/CMPB/plank2021,
  title={The openCARP simulation environment for cardiac electrophysiology},
  author={Plank, Gernot and Loewe, Axel and Neic, Aurel and Augustin, Christoph and Huang, Yung-Lin and Gsell, Matthias AF and Karabelas, Elias and Nothstein, Mark and Prassl, Anton J and S{\'a}nchez, Jorge and others},
  journal={Computer methods and Programs in Biomedicine},
  volume={208},
  pages={106223},
  year={2021},
  publisher={Elsevier}
}

@article{journal/MedIA/camps2025,
  title={Harnessing 12-lead ECG and MRI data to personalise repolarisation profiles in cardiac digital twin models for enhanced virtual drug testing},
  author={Camps, Julia and Wang, Zhinuo Jenny and Doste, Ruben and Berg, Lucas Arantes and Holmes, Maxx and Lawson, Brodie and Tomek, Jakub and Burrage, Kevin and Bueno-Orovio, Alfonso and Rodriguez, Blanca},
  journal={Medical image analysis},
  volume={100},
  pages={103361},
  year={2025},
  publisher={Elsevier}
}

@article{journal/brandstetter2022,
  title={Message passing neural PDE solvers},
  author={Brandstetter, Johannes and Worrall, Daniel and Welling, Max},
  journal={arXiv preprint arXiv:2202.03376},
  year={2022}
}

@article{journal/ANIPS/krishnapriyan2021,
  title={Characterizing possible failure modes in physics-informed neural networks},
  author={Krishnapriyan, Aditi and Gholami, Amir and Zhe, Shandian and Kirby, Robert and Mahoney, Michael},
  journal={Advances in neural information processing systems},
  volume={34},
  pages={26548--26560},
  year={2021}
}

@article{journal/FiP/minchole2019,
  title={MRI-based computational torso/biventricular multiscale models to investigate the impact of anatomical variability on the ECG QRS complex},
  author={Minchol{\'e}, Ana and Zacur, Ernesto and Ariga, Rina and Grau, Vicente and Rodriguez, Blanca},
  journal={Frontiers in physiology},
  volume={10},
  pages={1103},
  year={2019},
  publisher={Frontiers Media SA}
}

@article{journal/doste2019,
  title={A rule-based method to model myocardial fiber orientation in cardiac biventricular geometries with outflow tracts},
  author={Doste, Ruben and Soto-Iglesias, David and Bernardino, Gabriel and Alcaine, Alejandro and Sebastian, Rafael and Giffard-Roisin, Sophie and Sermesant, Maxime and Berruezo, Antonio and Sanchez-Quintana, Damian and Camara, Oscar},
  journal={International journal for numerical methods in biomedical engineering},
  volume={35},
  number={4},
  pages={e3185},
  year={2019},
  publisher={Wiley Online Library}
}

@article{journal/ANIPS/gruber2023,
  title={Reversible and irreversible bracket-based dynamics for deep graph neural networks},
  author={Gruber, Anthony and Lee, Kookjin and Trask, Nathaniel},
  journal={Advances in Neural Information Processing Systems},
  volume={36},
  pages={38454--38484},
  year={2023}
}

@inproceedings{conf/ICML/hansen2023,
  title={Learning physical models that can respect conservation laws},
  author={Hansen, Derek and Maddix, Danielle C and Alizadeh, Shima and Gupta, Gaurav and Mahoney, Michael W},
  booktitle={International Conference on Machine Learning},
  pages={12469--12510},
  year={2023},
  organization={PMLR}
}

@article{arxiv/baldan2026,
  title={GraphEP: A Calibrated Data System and Benchmark for Learning Cardiac Electrophysiology},
  author={Baldan, Matteo and D'Ercoli, Chiara and Rowbottom, James and Li{\`o}, Pietro and Fresca, Stefania},
  journal={arXiv preprint arXiv:2611.26551},
  year={2026}
}

@inproceedings{conf/CinC/costa2013,
  title={Automatic parameterization strategy for cardiac electrophysiology simulations},
  author={Costa, Caroline Mendonca and Hoetzl, Elena and Rocha, Bernardo Martins and Prassl, Anton J and Plank, Gernot},
  booktitle={Computing in Cardiology},
  pages={373--376},
  year={2013},
  organization={IEEE}
}

\end{document}